%% file: main.tex
\documentclass[pdflatex,sn-mathphys-ay]{sn-jnl}

\setcitestyle{authoryear,round,aysep={,},yysep={;},notesep={, }}
\usepackage{url}
\usepackage{graphicx,booktabs,amsmath,amssymb}
\usepackage[table]{xcolor}
\definecolor{hsscband}{RGB}{228,226,242}
\usepackage{caption}
\DeclareCaptionFormat{hssctab}{\colorbox{hsscband}{\parbox{\dimexpr\linewidth-2\fboxsep\relax}{\textbf{#1#2}#3}}}
\usepackage{algorithm}
\usepackage{algpseudocode}
\begin{document}
\title[Benchmarking agent societies]{Benchmarking large language model agent societies against human behavioural distributions}
\author*[1]{\fnm{Raad} \sur{Bin Tareaf}}\email{r.bintareaf@xu-university.de}
\affil*[1]{\orgdiv{AI Cluster}, \orgname{XU Exponential University of Applied Sciences}, \orgaddress{\street{Marlene-Dietrich-Allee 12B}, \postcode{14482}, \city{Potsdam}, \country{Germany}}}
\abstract{Populations of large language model agents are increasingly used as experimental
societies. Three doubts shadow every such result: whether the agents behave like the humans they
stand in for, whether a finding survives changes to the apparatus that leave the rules untouched,
and whether apparent social dynamics are interaction at all rather than the reproduction of
experiments the models have read. This article introduces SILICA, an open instrument that tests all
three. Five environments carry published human anchors, each paired with perturbations that
re-render the same rules and with variants whose payoffs point away from the memorised result.
Twelve open-weight models were run through it on a single consumer graphics card. Agreement with
human data is confined to starting points: first-round public-goods contributions fall inside the
equivalence margin for eight of eleven models, while no model matches end-state contributions or the
human corridor of cooperation. Merely swapping the order in which two actions are listed costs one
model 58 points of cooperation. Presenting responders with a fixed schedule of offers shows that
only one model, the sole reasoning-trained one, places its acceptance threshold where the incentive
requires; two move theirs part of the way, two move them the wrong way, and three never acquire one.
Conventions form through a shared prior over the names rather than through negotiation, though
negotiation reappears once that prior is disrupted. On the certification ladder defined here,
current silicon societies support exploratory claims and no more.}
\maketitle
\input{sections/intro_related_discussion_part1}

\input{sections/methods_merged}
\input{sections/setup_results}
\input{sections/intro_related_discussion_part2}

\backmatter
\bmhead{Acknowledgements}
The author acknowledges the use of an AI assistant (Anthropic Claude) for coding support and manuscript drafting under the author's direction; all experimental results were produced and verified by the author's infrastructure, and the author takes full responsibility for the content.
\section*{Declarations}
\textbf{Funding} No external funding was received. \\
\textbf{Competing interests} The author declares no competing interests. \\
\textbf{Ethics approval} Not applicable. This study did not involve human participants or animals. Human comparison data derive exclusively from published, publicly available datasets. \\
\textbf{Study plan and its timing} The design and analysis plan, the human anchor values and the analysis code were deposited publicly on the Open Science Framework at 07:02~UTC on 15~August~2026 and have not been modified since (\url{https://doi.org/10.17605/OSF.IO/AFSDU}); the same material is tagged \texttt{prereg-v1} in the code repository, committed at 06:53~UTC that day. The first run in the released record began at 10:05~UTC, about three hours later. The deposit was not converted into a frozen OSF registration, so it is described here as a timestamped public deposit rather than as a registration. Departures from the deposited plan are disclosed in the Methods and in the Supplementary Information. \\
\textbf{Data availability} The complete run-level record (9{,}115 model runs and 150 runs of the classical baselines, including the offer-grid and shuffled-pool arms), every aggregated statistic, the full test-level record with exact $p$-values, the provenance of each human anchor, and the complete transcript corpus (8{,}220 files, one per run) are openly available. The transcripts and the code archive are deposited at \url{https://doi.org/10.5281/zenodo.22091078}; the numbers reported here come from version 1.1.0 of that record (\url{https://doi.org/10.5281/zenodo.22122939}). The human comparison data are not redistributed here: they remain with their original publishers and are cited in the anchor provenance table. \\
\textbf{Code availability} The complete benchmark --- environments, perturbation library, contamination generator, execution harness, and every analysis and figure script --- is released under the MIT licence at \url{https://github.com/raadbintareaf/silica-benchmark}. Release \texttt{v1.1.0} is the version that reproduces every number reported here from the released run record; earlier tags in the repository predate this analysis and the differences are listed in the release changelog. \\
\textbf{Author contributions} R.B.T. conceived the study, designed the instrument and its environments, implemented the benchmark and the analysis pipeline, ran all experiments, carried out the analysis, produced the figures and tables, and wrote the article.
\bibliography{references}
\end{document}


\maketitle

\section*{Contents}
\begin{figure*}[htbp]\centering
\includegraphics[width=\textwidth]{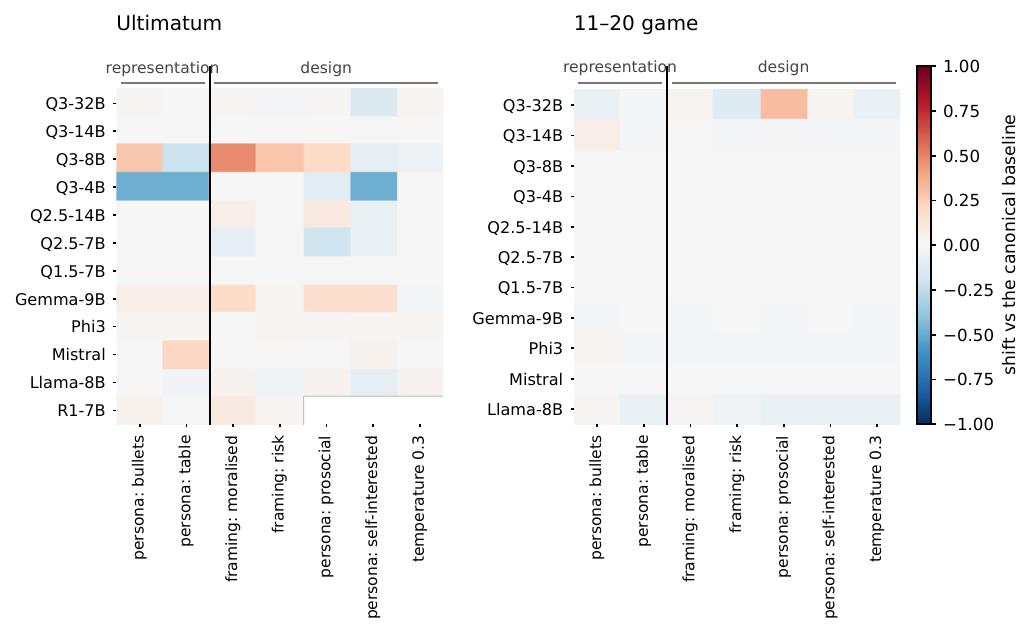}
\caption{Perturbation cells in the two one-shot environments. Rows are models, columns are cells,
ordered representation-level first. These cells hold two runs each, so the shifts are descriptive
and no significance is marked; the three environments in which run-level tests are possible appear
in the main text.}
\label{fig:si-oneshot}
\end{figure*}

\begin{figure*}[htbp]\centering
\includegraphics[width=\textwidth]{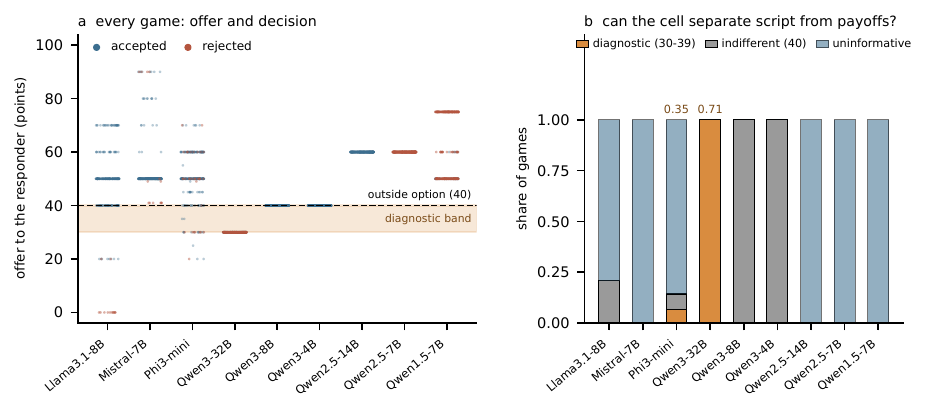}
\caption{The divergent-payoff ultimatum cell as originally run, before the fixed offer schedule was
introduced. \textbf{a} Every game, plotted as the offer against the responder's decision; the shaded
band marks the range in which the memorised script and the payoffs disagree. \textbf{b} The share of
each model's games that fall in that band. Seven of nine models never produce a single diagnostic
game, because the proposer is the same model as the responder and offers at or above the outside
option. This is the observation that motivated the fixed schedule reported in the main text.}
\label{fig:si-ultimatum}
\end{figure*}

\begin{figure*}[htbp]\centering
\includegraphics[width=\textwidth]{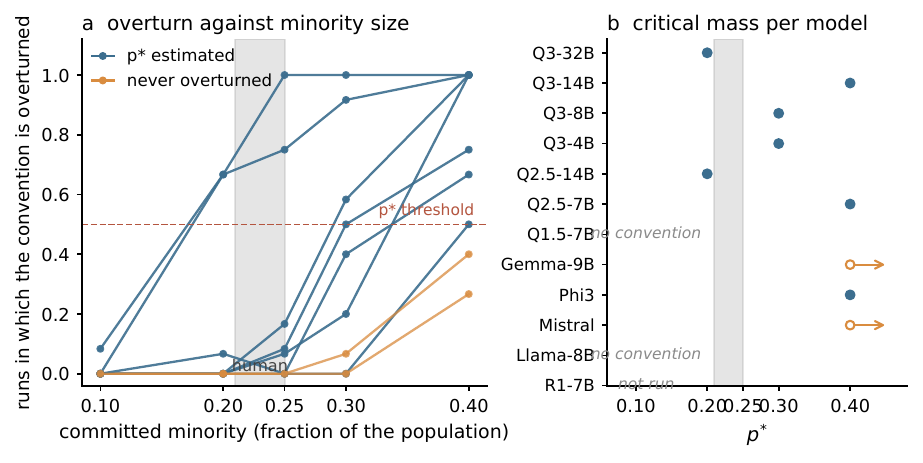}
\caption{Committed-minority tipping, recomputed. \textbf{a} Share of runs in which an established
convention is overturned, against the size of the committed minority; the shaded band is the human
bracket, 21--25\% of the group. Lines are blue where a critical mass could be estimated and orange
where the convention was never overturned at any tested fraction. \textbf{b} The resulting $p^{*}$
per model, defined as the smallest tested fraction at which the convention is overturned in at least
half of runs. Open circles with an arrow mark models whose conventions survive every tested minority;
models that never form a convention have nothing to overturn, and the tipping arm was not executed
for R1-Distill.}
\label{fig:si-tipping}
\end{figure*}

\begin{figure}[htbp]\centering
\includegraphics[width=\linewidth]{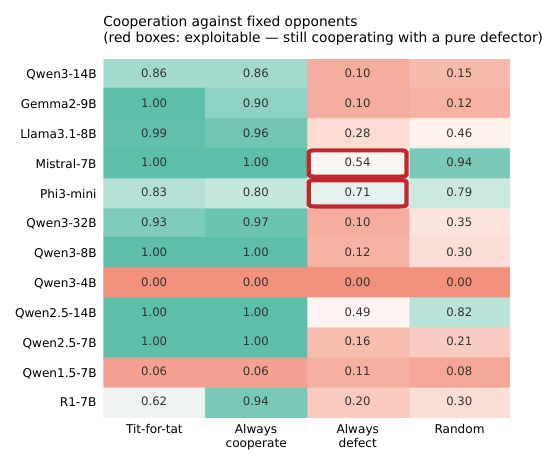}
\caption{Single-agent diagnostics: cooperation against four fixed opponents. These runs are
diagnostics rather than perturbation cells, and they were excluded from the baseline in the revised
analysis; the figure is unaffected by that correction because it plots the diagnostic runs
themselves. Two models cooperate with an opponent that always defects, which is the clearest single
indication that behaviour is not payoff-driven.}
\label{fig:si-single}
\end{figure}

\begin{figure*}[htbp]\centering
\includegraphics[width=\textwidth]{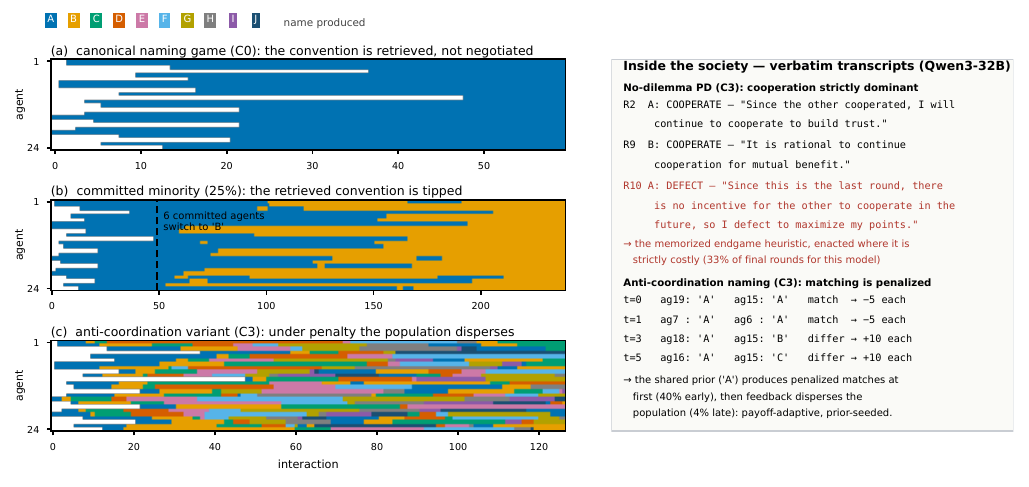}
\caption{A society in progress: one run rendered as a raster of agents against rounds, with an
excerpt of the recorded transcript. Included because the corrections reported in the main text were
found by reading transcripts against outcome statistics, and this is what that reading looks like.}
\label{fig:si-society}
\end{figure*}

This file supports the main article. It contains the model roster and execution statistics
(Table~\ref{tab:si-roster}), the test-level record for every contrast named in the main text
(Table~\ref{tab:si-stats}), contamination outcomes away from the divergent-payoff variant with
recognition-probe rates (Table~\ref{tab:si-contam}), the provenance of every human anchor
(Table~\ref{tab:si-anchors}), the acceptance functions under the fixed offer schedule
(Table~\ref{tab:si-grid}), the shuffled-pool results (Table~\ref{tab:si-shufpool}), the positioning of this article against the closest prior work (Table~\ref{tab:positioning}), and
Supplementary Figures~\ref{fig:si-oneshot}--\ref{fig:si-society}. All values are recomputed from the released run record by the released scripts; nothing here is
transcribed by hand. The record is archived at \url{https://doi.org/10.5281/zenodo.22091078},
version 1.1.0.

\section*{Supplementary tables}
\input{tables/si_roster}
\input{tables/si_stats}
\input{tables/si_contam}
\input{tables/si_anchors}
\input{tables/si_grid}
\input{tables/si_shufpool}

\subsection*{Criteria used in the positioning table}
Each column of Table~\ref{tab:positioning} applies one criterion, stated here so that a reader can
check the marks rather than take them on trust.
\emph{Interactive multi-agent}: agents act on one another's behaviour within a run, rather than
being sampled independently.
\emph{Human-anchored (distribution)}: outcomes are compared with a published human quantity, and
the comparison is to a distribution or a stated interval rather than to a single illustrative value.
\emph{Perturbation audit}: the same rules are re-rendered, or the content surrounding them altered,
and the sensitivity of the result is reported.
\emph{Contamination audit (C3)}: a divergent-prediction design is \emph{executed} --- payoffs
altered so that the memorised result is no longer the profitable one --- rather than contamination
being obfuscated, discussed, or tested only by recognition probes.
\emph{Five or more model families}: results reported for models from at least five distinct
developers, the point at which a claim about ``language-model agents'' becomes more than a claim
about one lineage.
\emph{Scale axis}: at least three sizes within one family.
\emph{Generation axis}: at least two releases from one developer at comparable size.
\emph{Claim certification}: findings are graded by an explicit evidentiary standard rather than
reported uniformly.
\emph{Open executable benchmark}: environments, configurations and analysis code are released in a
form that re-runs the study, as opposed to code released for inspection only.

\input{tables/positioning}

\clearpage
\section*{Supplementary figures}

\clearpage
\bibliography{references}

%% file: sections/intro_related_discussion_part1.tex
\section{Introduction}\label{sec:intro}

Populations of large language model (LLM) agents are rapidly becoming instruments of social science. Generative agents form relationships and plan parties \citep{park2023}, interview-grounded agents reproduce a thousand people's survey answers \citep{park2024}, urban simulators run policy experiments over tens of thousands of agents \citep{piao2025agentsociety,yang2024oasis}, and decentralized LLM populations appear to develop shared conventions and tipping points that mirror human collective dynamics \citep{ashery2025}. The promise is a laboratory without recruitment costs \citep{horton2023,anthis2025}, and the paradigm has been mapped in comprehensive surveys of agent-based modelling with language models \citep{gao2024hssc,mou2026}; most recently, an ``AI agent behavioural science'' has been proposed that treats agent behaviour --- rather than model internals --- as the object of systematic observation and intervention \citep{chen2026hssc}. What that program lacks for \emph{social} behaviour is a calibrated measurement instrument; supplying one is this paper's purpose.

The validity of that laboratory, however, is contested on three fronts, so far mostly in position papers. First, \emph{fidelity}: replications report inflated effect sizes and unpredictable distribution-level failures \citep{cui2025,gao2025pnas}, and public-opinion simulations exhibit systematic demographic biases \citep{quwang2024}, and even the strongest agent architectures degrade precisely in interactive economic games \citep{park2024}. Second, \emph{robustness}: semantically irrelevant details of the apparatus --- how a persona is formatted --- can move outcomes by tens of percentage points \citep{ye2026}, and sensitivity analysis is rarely performed \citep{larooij2025,zhou2025pimmur}. Third, \emph{contamination}: the flagship emergence results are observationally equivalent to reproduction of memorised experimental scripts, a critique that existing mitigations do not answer \citep{barrie2025,ashery2025reply}. What the debate lacks is an instrument: prior benchmarks evaluate either static, single-turn response distributions \citep{hu2025simbench} or model \emph{capability} in games \citep{duan2024gtbench,shapira2024glee}, and the robustness and contamination programs exist as taxonomies and proposals \citep{ye2026,payne2026} rather than executed measurements. No released system evaluates \emph{interactive multi-agent} simulations against human ground truth under systematic perturbation and contamination audits.

This paper builds that instrument and reports what it finds. The contributions are:

\begin{enumerate}
\item \textbf{SILICA}, an open instrument that couples versioned, seeded, interactive multi-agent
environments --- a repeated prisoner's dilemma, public goods with costly punishment, a bargaining
battery, the 11--20 money-request game, and a naming game with committed-minority tipping --- with
published human anchors whose provenance is released as data rather than asserted as constants.
\item An \textbf{executable perturbation library} that separates changes which re-render the same
information from changes which alter it. Design-level changes move behaviour in 56 of 111 computable
contrasts and representation-level changes in 8 of 71 --- so most fragility is a response to content,
but a minority is not: rendering a persona as a table rather than a sentence ends convention
formation in one model, and swapping the order in which two actions are listed costs another 58
points of cooperation.
\item A \textbf{contamination audit} whose central instrument is a fixed schedule of offers, which
identifies each model's acceptance function independently of what its own proposers happen to offer.
One model of twelve places that threshold where the outside
option puts it; two move theirs part of the way, two move them the wrong way, three never
acquire one, and one acquires a boundary only when the payoffs change and then puts it at the
human value rather than the one the incentive specifies. No aggregate rejection rate
distinguishes these cases.
\item A \textbf{control that overturned this study's own mechanism claim}. Conventions appear to form
without negotiation --- until each agent is shown its own permutation of the same name pool, after
which coordination survives but takes one and a half to two and a half times longer in seven
models of eight, and stops being reliable in the eighth. The finding was real and
its explanation was not, and the manipulation that separated them was absent from the perturbation
library until a transcript audit suggested it.
\item A \textbf{tiered certification protocol} whose application yields a measured evidentiary
ceiling: across 9{,}115 runs of twelve models spanning family, scale, generation and
reasoning-training axes, no finding is transferable, convention formation alone is robust, and every
other finding is exploratory.
\end{enumerate}

Section~\ref{sec:related} situates the work, Section~\ref{sec:methods} specifies the instrument and the statistical protocol, Section~\ref{sec:results} reports the findings, and Section~\ref{sec:discussion} sets out what they permit and what they do not.

The instrument is organised around four questions, and the article answers them in order.
\textbf{RQ1, fidelity:} do populations of language-model agents reproduce the levels and the
directions of behaviour recorded in human experiments, judged by equivalence tests against
published anchors rather than by visual similarity? \textbf{RQ2, robustness:} does a finding
survive changes to the apparatus that leave the rules of the game untouched, and does it matter
whether those changes re-render the same information or alter the content the agents are given?
\textbf{RQ3, provenance:} when the payoffs of a familiar experiment are altered so that the
memorised result is no longer the profitable one, do the agents follow the payoffs, follow the
memorised script, or fail to respond to the situation at all? \textbf{RQ4, certification:} taken
together, what standard of claim can a study built on these societies currently support?

\section{Related work}\label{sec:related}

\textbf{What human experiments established, and how.} The behaviours this article uses as anchors
were not established by single studies but by programmes of experiments whose designs are worth
recalling, because they set the standard an agent society has to meet. Conventions were shown to
emerge from purely local interaction, without central coordination and without any participant
seeing the population, in laboratory naming games \citep{centola2015}; the same paradigm later
established that a committed minority can overturn an established convention once it passes a
threshold, with the transition observed between 21 and 25\% of the group \citep{centola2018}.
Cooperation in repeated dilemmas, contribution decay in public goods, and the effect of costly
punishment are each supported by dozens of replications summarised in meta-analyses
\citep{mengel2018,oosterbeek2004,engel2011,johnson2011}, and the theoretical account of how norms
spread and stabilise is correspondingly well developed \citep{young2015}. Two features of that
literature matter here. Its central results are distributions and trajectories rather than point
estimates, which is why this article tests them with equivalence rather than similarity. And its
mechanisms are established by manipulation --- varying the minority size, the punishment
technology, the information each participant sees --- rather than by observing that a simulated
population produced a similar-looking outcome.

\textbf{Validation as a discipline, not a step.} Agent-based modelling arrived at this problem
decades ago and produced a literature on what it takes to believe a simulated population: which
empirical targets a model must match, at what level of aggregation, and what remains unidentified
when several parameterisations reproduce the same output \citep{fagiolo2007}. The practical
guidance that followed --- state the target, state the tolerance, report the sensitivity of the
result to modelling choices that are not part of the theory \citep{rand2011} --- maps directly onto
what an instrument for language-model societies needs, and this article's perturbation library is
that sensitivity analysis under a different name. The independent-verification movement in the
experimental social sciences supplies the second half of the standard: large-scale replication
projects put numbers on how often published effects survive re-testing
\citep{osc2015,camerer2016,camerer2018}, and pre-registration became the routine defence against
analytic flexibility \citep{nosek2018}. An instrument that reports only that a simulated population
behaved plausibly meets neither standard.

\textbf{Generative social simulation.} Sandbox societies \citep{park2023,vezhnevets2023},
large-scale simulators \citep{piao2025agentsociety,yang2024oasis}, and individual-level ``silicon
sampling'' \citep{argyle2023,horton2023,aher2023,santurkar2023} established the paradigm;
interview-grounded agents mark its highest reported fidelity \citep{park2024}. Emergence-flavoured
findings include conventions and tipping \citep{ashery2025} and norm formation \citep{ren2024}.
This article differs in treating such claims as hypotheses to be tested against human anchors
rather than as demonstrations.

\textbf{Games as evaluation, and the validation turn.} A growing set of benchmarks measures
strategic \emph{capability} --- matrix and extensive-form suites, economic environments,
behavioural-game-theoretic depth, and process-aware mixed-motive evaluation
\citep{duan2024gtbench,huang2024gama,tmgbench2024,shapira2024glee,klevel2025,m3bench2026} --- and
repeated-game work reports behavioural signatures that diverge from human ones
\citep{akata2025,lore2024,fontana2025}. These instruments rank models; none anchors interactive
populations to human distributions or audits validity. Alongside them, critical reviews argue that
validation is the central unresolved problem \citep{larooij2025,mou2026}, replication programmes
quantify vignette-level fidelity and its inflation \citep{cui2025}, distribution-level failures
\citep{gao2025pnas} and identity flattening \citep{wang2025} sharpen the caution, and
methodological principles, taxonomies and infrastructure continue to accumulate
\citep{zhou2025pimmur,boundary2025,lin2026,anthis2025,ye2026,hu2025simbench,sarangi2026}.
Supplementary Table~S7 positions this article against the closest of them: the vacancy it occupies
is the intersection --- interactive, multi-agent, human-anchored, perturbation-audited,
contamination-controlled.

\textbf{Contamination.} Detection methods target static benchmarks. For interactive experiments the
dispute over the naming game \citep{barrie2025,ashery2025reply} and a proposed novel-game
falsification design \citep{payne2026} remained untested. The audit reported here executes that
design and extends it to collective dynamics, with recognition probes and priming bounds in the
spirit of \citet{gao2025pnas}.

%% file: sections/methods_merged.tex
\section{Methods}\label{sec:methods}\label{sec:method}

\begin{sidewaysfigure}[p]
\centering
\includegraphics[width=\textheight]{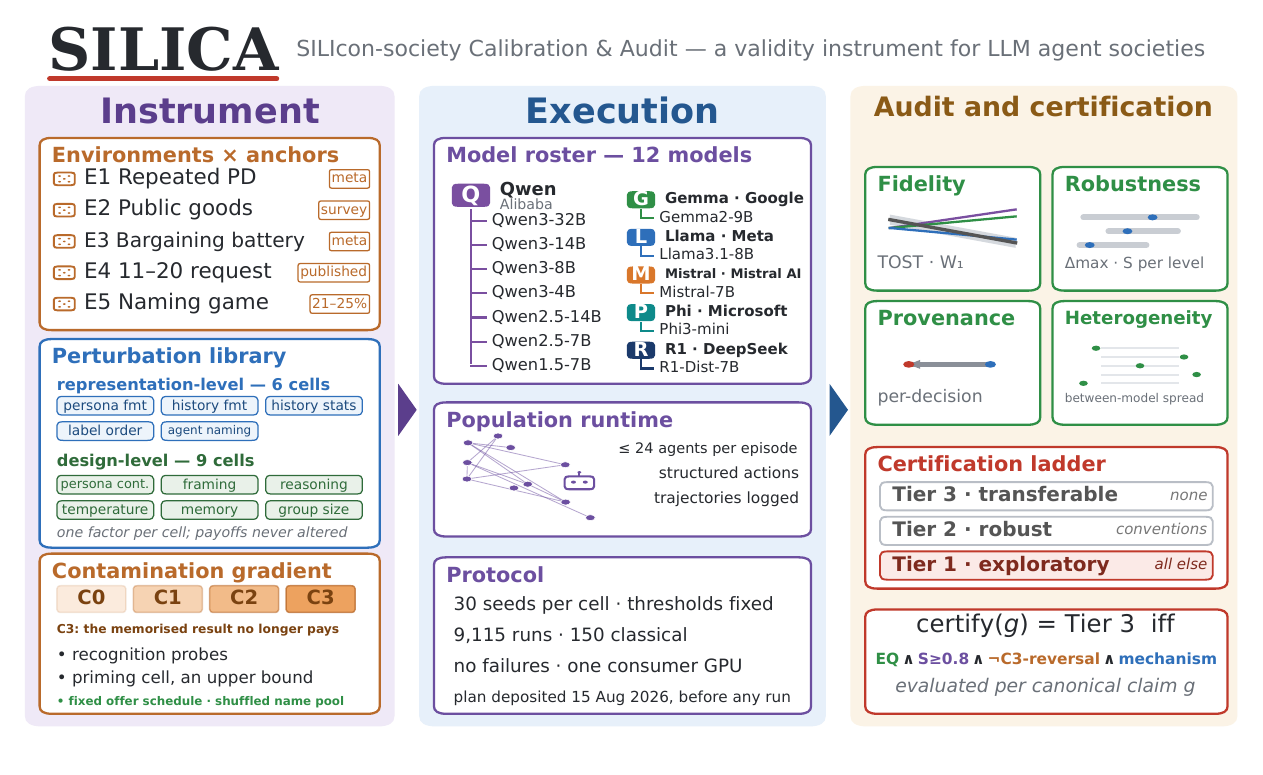}
\caption{The instrument. \textbf{Left} Five environments, each paired with a published human
anchor, a perturbation library whose cells are separated into those that re-render the same rules
and those that alter what the agents are told, and a gradient of payoff variants ending in C3,
where the result recorded in the literature is no longer the profitable one; the last two entries
are controls added after a transcript audit. \textbf{Centre} The roster and the execution protocol; the full roster with quantisation and per-model run counts is in the Supplementary Information.
\textbf{Right} The four measurements and the ladder that converts them into a claim grade, with the
current verdict shown against each tier. Cell counts and run totals are read from the released
record; the registration shown in the protocol card is at
\url{https://doi.org/10.17605/OSF.IO/AFSDU}.}
\label{fig:instrument}
\end{sidewaysfigure}

\subsection{Environments and their human anchors}\label{sec:envs}
Figure~\ref{fig:instrument} sets out the instrument as a whole; the released implementation is \citet{silica2026software}.
Five environments were selected because each has a well-documented human result that can be stated
as a number rather than as a description, and because between them they span the forms of social
behaviour that agent societies are usually asked to reproduce: cooperation under temptation,
contribution to a shared pool, bargaining over a fixed sum, iterated strategic reasoning, and the
emergence of a shared convention.

\emph{Repeated prisoner's dilemma.} Two agents choose simultaneously for ten rounds with the
canonical payoffs $(R,S,T,P) = (3,0,5,1)$. The anchor is the corridor of average cooperation rates
reported across prisoner's dilemma studies \citep{mengel2018}, together with the decline of cooperation over rounds
and a rise in defection in the final round.

\emph{Public goods with punishment.} Four agents each receive an endowment of 20 points per round
for ten rounds and choose a contribution to a common pool with a marginal per-capita return of
$0.4$. In the punishment arm, each agent may afterwards spend up to 10 points per target at a cost
of 1 point per 3 points removed. The anchors are a first-round contribution near half the endowment
and a decline over repetition, together with the conditional-cooperation pattern that underlies it \citep{fischbacher2001}; the punishment anchor comes from a single experiment \citep{fehr2000} and is labelled as such.

\emph{Bargaining battery.} Ultimatum, dictator and trust games, 50 one-shot games per run, with a
pie of 100 points in the ultimatum game. Anchors are meta-analytic: a mean offer of $0.40$ of the
pie and an overall rejection rate of $0.16$; dictator giving of $0.28$ of endowment; and trust
sending of $0.50$ with $0.37$ of the available amount returned.

\emph{The 11--20 money-request game.} Each agent requests an integer between 11 and 20 and receives
it, plus a bonus of 20 if the request is exactly one below the other agent's; 50 one-shot games per
run. The anchor is the full choice distribution reported by \citet{arad2012}, whose mode is 17.

\emph{Naming game with tipping.} Twenty-four agents are paired at random and each proposes a name
from a shared pool of ten, with a memory of the last five interactions; a convention is recorded
when at least 95\% of the last 50 interactions match, and a run is capped at 3{,}000 interactions.
A separate arm introduces a committed minority of between 10 and 40\% of the population. The
anchors are that a convention always forms \citep{centola2015}, and that the critical mass sits between 21 and 25\% of the population \citep{centola2018}.

Every anchor value, its source, the size of the evidence base behind it, and its provenance tier is
given in Supplementary Table~S4, together with the correction made to the 11--20 anchor. One anchor was corrected during this work: two cells of the 11--20
distribution had been mis-transcribed, and all distances are computed against the published table.

\subsection{The perturbation library}\label{sec:perturb}
A perturbation changes the apparatus without changing the game. Each cell alters exactly one factor
and holds everything else at the canonical baseline (plain persona, neutral content, canonical
framing, tabular history without summary statistics, default action-label order, letter agent
names, temperature $0.7$, memory window 5, default group size). Table~\ref{tab:cells} lists every
cell and the environments in which it ran.

\input{tables/cell_list}

Cells are of two kinds, and the distinction matters for what a shift means.
\emph{Representation-level} cells re-render information the agents already have: the persona as
bullets or as a table, the history as narrative prose, a summary of statistics the history already
contains, the two actions listed in the opposite order, agents given names instead of letters. A
model whose behaviour changes here has changed it for no informational reason at all.
\emph{Design-level} cells change what the agents are given: a self-interested or prosocial persona,
a moralised framing that relabels the actions as acting fairly and exploiting, a risk framing that
labels them safer and riskier, an instruction to reason step by step before answering, a lower
sampling temperature, a shorter memory window, a different group size. Framing sits with the design
cells because it adds evaluative content rather than re-rendering existing content.

\subsection{Variants that separate memory from computation}\label{sec:contam}
Four versions of each environment were run. C0 is the canonical form. C1 re-skins it with a
different cover story and different action labels while preserving the payoffs. C2 changes the
surface and shifts the numbers. C3 changes the payoffs so that the result recorded in the human
literature is no longer the profitable one: cooperation becomes strictly dominant in the repeated
dilemma; the naming game rewards mismatching rather than matching; the responder in the ultimatum
game receives a fixed 40 points on rejection; and the 11--20 bonus is paid for exceeding rather
than undercutting the other player. Two further probes complete the audit: a recognition probe,
asked outside the behavioural session, records whether a model can name the paradigm from the
instruction text, and a priming cell names the paradigm inside the behavioural prompt.

Because the proposer and the responder in the ultimatum game are the same model, whether a model
ever faces an offer in the range where memory and payoffs disagree is itself an outcome. A
supplementary arm therefore presents responders with a fixed schedule of offers --- 10, 20, 25, 30,
33, 37, 40, 45 and 50 points, twenty games each --- in both the canonical and the divergent-payoff
version, so that the acceptance function is identified for every model on the same decisions (Supplementary Table~S5).
Behaviour under the divergent-payoff variant is classified as following the payoffs, following the
script, or mixed,

\begin{equation}\label{eq:c3}
\mathrm{C3}(E,m) \;=\;
\begin{cases}
\textsf{follows-payoffs} & v \geq \tau^{\ast},\\
\textsf{follows-script} & v \leq \tau^{s},\\
\textsf{mixed} & \text{otherwise},
\end{cases}
\end{equation}
where $v$ is the share of decisions consistent with the altered payoffs. In the ultimatum game that
share is evaluated per game against the offer actually faced, because rejection pays only when the
offer falls below the outside option; an aggregate rejection rate cannot express this rule and the
thresholds $\tau^{\ast} = 0.70$ and $\tau^{s} = 0.30$ apply to the per-game share.

A second supplementary arm addresses the naming game: because the ten names are otherwise presented
in the same order to every agent, each agent receives its own random permutation of the pool, and
an unparsable reply falls back to a random name rather than to the first listed one (Supplementary Table~S6).

\subsection{Models, serving and decoding}\label{sec:models}
Twelve open-weight models were served locally with vLLM \citep{kwon2023vllm} through an OpenAI-compatible interface on a single consumer graphics card. Quantisation varies across the roster and is recorded per model in
Supplementary Table~S1: three models are served in AWQ, one in FP8, and the remaining eight at
native precision.
Sampling temperature is $0.7$ except in the temperature cell. The Qwen3 models expose an optional
deliberation mode, which was disabled throughout so that the reasoning axis is carried by the
single reasoning-trained model rather than by a decoding switch; where a model emits deliberation
markers, they are stripped before parsing. Agents return a single structured action per call
against a schema fixed by the environment; an optional free-text reason is recorded but never
parsed into behaviour.

\subsection{Run structure}\label{sec:runs}\label{sec:setup}
Each repeated-game cell holds 30 seeded runs per model at baseline and 10 in each perturbation
cell. Each one-shot cell holds 6 runs of 50 games at baseline and 2 runs of 50 games in each
perturbation cell; results from the latter are reported as descriptive throughout, because two runs
cannot support an inferential test. Seeds control agent order, pairing and sampling. These counts hold for every model except R1-Distill,
which completed a partial battery: 11 baseline runs in the repeated dilemma, three in each one-shot
baseline, and between one and seven in the seventeen perturbation cells it reached. Every table
entry that would rest on fewer than three runs is shown as absent rather than as a value. The
complete matrix comprises 9{,}115 model runs and 150 runs of the classical baselines, with no
execution failures.

\subsection{Outcomes and how they are compared}\label{sec:outcomes}
Each environment has one primary outcome: the cooperation rate; the contribution as a fraction of
endowment in rounds 1 and 10; the offer as a fraction of the pie; the share of choices at the human
mode; and whether a convention forms. Agreement with an anchor is assessed by two one-sided tests
at a margin fixed in advance and deposited before the first run --- $\pm0.05$ for ultimatum offers and dictator giving, $\pm0.07$ for
the trust quantities, $\pm0.10$ for public-goods contributions --- and never by visual inspection.
The two one-sided tests at $\alpha = 0.05$ are equivalent to requiring that the 90\% confidence interval of the difference lie inside the margin, which is the form used here,

\begin{equation}\label{eq:tost}
\mathrm{EQ}(E,m) \;=\; \mathbf{1}\!\left[\, \mathrm{CI}_{90\%}\!\left(\bar{y}(E,m,c_0) - \theta_H(E)\right) \subseteq [-\varepsilon_E, \varepsilon_E] \,\right],
\end{equation}
where $\bar{y}(E,m,c_0)$ is the model's mean at the canonical baseline, $\theta_H(E)$ the human
anchor and $\varepsilon_E$ the margin fixed in advance.

For the 11--20 game, whose anchor is a whole distribution, agreement is measured by the Wasserstein
distance between the pooled choice histogram and the published one, normalised by the width of the
support, with a bootstrap interval over runs,

\begin{equation}\label{eq:w1}
\widetilde{W}_1(E,m) \;=\; \frac{W_1\!\left(D_L(E,m,c_0),\, D_H(E)\right)}{\max \Omega_E - \min \Omega_E},
\end{equation}
with $D_L$ the pooled choice histogram, $D_H$ the published distribution and $\Omega_E$ the support.

Two further quantities summarise the perturbation library. $\Delta_{\max}$ is the range of cell
means over the baseline and the cells of one level, and $S$ is the share of those cells in which
the canonical human effect keeps its sign. Both are reported separately for representation-level
and design-level cells,

\begin{equation}\label{eq:swing}
\Delta_{\max}(E,m) \;=\; \max_{c \in \mathcal{C}_E} \bar{y}_c \;-\; \min_{c \in \mathcal{C}_E} \bar{y}_c,
\qquad
S(E,m) \;=\; \frac{1}{|\mathcal{C}_E|} \sum_{c \in \mathcal{C}_E} \mathbf{1}\!\left[\operatorname{sign} g(E,m,c) = \sigma_H \right],
\end{equation}
where $\mathcal{C}_E$ ranges over the baseline and the cells of one level, and $\sigma_H$ is the
sign of the canonical human effect.

Directional agreement uses one rule in every environment: the median over a model's baseline runs of
the run-level effect must clear a margin of $0.10$ on the outcome's own scale, where the effect is
the change in contributions between the first and last round, the modelled change in cooperation
across the ten rounds, or, for a rate outcome, formation of a convention in the majority of runs.

\subsection{Statistical procedure}\label{sec:stats}
Comparisons between a cell and its baseline use the Mann--Whitney rank-sum test, evaluated against
the exact permutation distribution of the statistic conditional on the observed ties rather than
against a normal approximation, so that reported values never fall below the smallest attainable
$p$ for the sample sizes involved. Effect sizes are Cliff's $\delta$. Within each family of tests
--- perturbation, priming, punishment, contamination --- $p$-values are adjusted by the Holm
step-down procedure, and a contrast is called significant when the adjusted value is below $0.05$.
Contrasts in which either arm holds fewer than three runs, or in which the pooled values are
constant, carry no evidence and are excluded from the family rather than entered with $p = 1$.
Between-model heterogeneity is reported as $I^{2}$ alongside the between-model standard deviation
of baseline means, because with 30 runs per model and small within-model variance $I^{2}$ saturates
whenever model means differ at all.

\subsection{Certification}\label{sec:tiers}
The instrument grades a claim rather than a model. A claim reaches \emph{Tier~1, exploratory}, if it
holds at baseline in at least one model. It reaches \emph{Tier~2, robust}, if it additionally
survives the perturbation library --- sign stability of at least $0.8$ across cells in at least
three model families --- and is not reversed under the divergent-payoff variant. It reaches
\emph{Tier~3, transferable}, only if it additionally matches the human anchor by the equivalence
test rather than merely in direction, and if the transcript audit is consistent with the mechanism
recorded in the human experiment. The ladder is a design choice of this instrument. It was not part of the deposited analysis plan; the equivalence margins, the primary outcomes and the contamination thresholds for the repeated dilemma, the 11--20 game and the naming game were. The design and analysis plan, the human anchor values and the analysis code were deposited publicly on the Open Science Framework at 07:02~UTC on 15~August~2026 and have not been modified since (\url{https://doi.org/10.17605/OSF.IO/AFSDU}); the same material is tagged \texttt{prereg-v1} in the code repository, committed at 06:53~UTC that day. The first run in the released record began at 10:05~UTC, about three hours later. The deposit was not converted into a frozen OSF registration, so it is described here as a timestamped public deposit rather than as a registration.

\subsection{Disclosure of language-model use}\label{sec:llmuse}
A large language model was used during the preparation of this article for code review, for
drafting and editing prose, and for cross-checking bibliographic records against Crossref and
arXiv. It was not used to generate data, to select analyses, or to interpret results. Every
numerical value reported here was recomputed from the released run record by the released scripts,
and every bibliographic record was confirmed against its primary source by the author.

%% file: tables/cell_list.tex
\begin{table}[htbp]
\caption{The perturbation library: every cell of the audit grid, its level, and the environments in which it was run ($\bullet$). One factor changes per cell; all other components are held at the canonical baseline (plain persona, neutral content, canonical framing, tabular history without summary statistics, default label order, letter labels, temperature $0.7$, memory window $5$, default group size). Each cell holds 10 seeded runs per model (2 runs of 50 games in the one-shot battery).}
\label{tab:cells}\scriptsize\setlength{\tabcolsep}{3pt}\centering
\begin{tabular}{@{}llccccc@{}}
\toprule
Level & Cell & PD & PGG & Ult. & 11--20 & Naming \\
\midrule
Representation & agent naming = names & $\bullet$ & $\bullet$ &  &  &  \\
 & history rendering = narrative & $\bullet$ &  &  &  &  \\
 & history summary statistics = shown & $\bullet$ & $\bullet$ &  &  &  \\
 & action-label order = swapped & $\bullet$ &  &  &  &  \\
 & persona layout = bullets & $\bullet$ & $\bullet$ & $\bullet$ & $\bullet$ & $\bullet$ \\
 & persona layout = table & $\bullet$ & $\bullet$ & $\bullet$ & $\bullet$ & $\bullet$ \\
\midrule
Design & step-by-step reasoning & $\bullet$ &  &  &  &  \\
 & framing = moralized & $\bullet$ & $\bullet$ & $\bullet$ & $\bullet$ &  \\
 & framing = risk & $\bullet$ & $\bullet$ & $\bullet$ & $\bullet$ &  \\
 & group size = 50 &  &  &  &  & $\bullet$ \\
 & group size = 8 &  & $\bullet$ &  &  &  \\
 & memory window = 2 &  &  &  &  & $\bullet$ \\
 & persona content = prosocial & $\bullet$ & $\bullet$ & $\bullet$ & $\bullet$ & $\bullet$ \\
 & persona content = self-interested & $\bullet$ & $\bullet$ & $\bullet$ & $\bullet$ & $\bullet$ \\
 & sampling temperature = 0.3 & $\bullet$ & $\bullet$ & $\bullet$ & $\bullet$ & $\bullet$ \\
\bottomrule
\end{tabular}
\end{table}

%% file: sections/setup_results.tex
\section{Results}\label{sec:results}

Table~\ref{tab:glance} condenses the study; the subsections unpack each column.

\input{tables/glance}

\subsection{Fidelity: agreement is confined to starting points}\label{sec:rq1}
Table~\ref{tab:fidelity} reports the equivalence tests, with the committed-minority results behind its $p^{*}$ column in Supplementary Fig.~S3; the test-level record for every contrast named below is Supplementary Table~S2. Agreement with the human anchors appears
where behaviour begins and disappears where it develops. First-round public-goods contributions
fall inside the margin fixed in advance for eight of the eleven models with a complete cell; by the
final round, none does. Ultimatum offers match the meta-analytic mean of 0.40 of the pie for one
model of twelve. Dictator giving matches for none, and no model's cooperation rate falls inside the
corridor of 0.20 to 0.50 reported across prisoner's dilemma studies --- every model is either far
more cooperative than the human range or entirely uncooperative.

Direction fares only slightly better. Under a single rule, applied identically in every environment
--- the median run-level change must clear a margin of 0.10 on the outcome's own scale --- the
canonical decline of public-goods contributions is reproduced by one model of twelve, the decline of
cooperation in the repeated dilemma by three, and convention formation by ten. Of the eleven models with a complete cell, seven end the public-goods game contributing
\emph{more} than they started and three end exactly where they began; only one declines.

The classical baselines make the comparison concrete. A conditional-cooperator model \citep{fischbacher2001} run in the same
environment declines from 0.48 to 0.07 of endowment, reproducing the human trajectory that ten of
eleven language models miss; a Roth--Erev learner \citep{roth1995} declines by only 0.03 and does not clear the
margin, and in the repeated dilemma that same learner cooperates in 0.42 of rounds, inside the human
corridor that no language model reaches.

\begin{figure*}[t]\centering
\includegraphics[width=\linewidth]{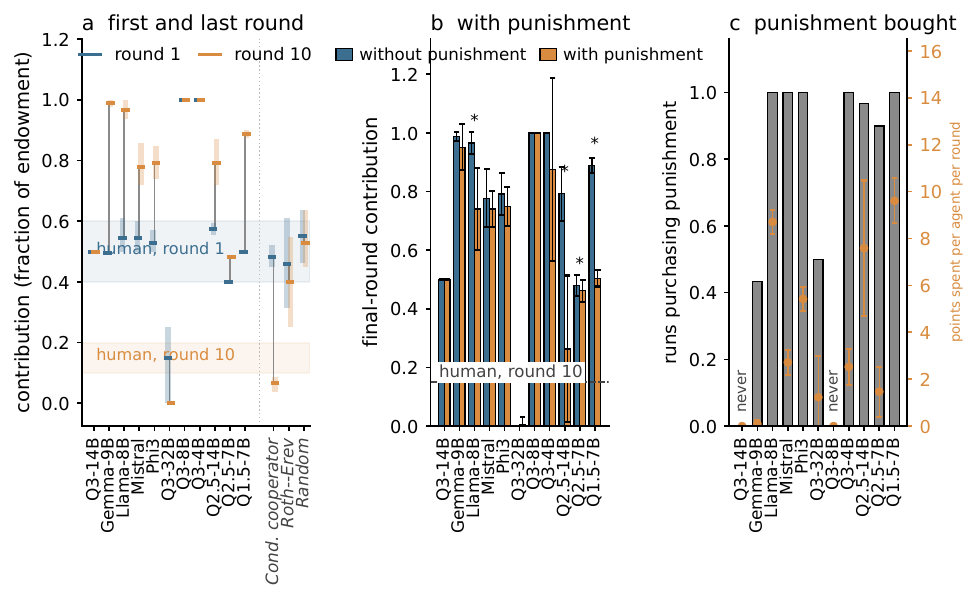}
\caption{Public goods and costly punishment. \textbf{a} First- and last-round contributions at the canonical baseline, with the interquartile range across runs and the human corridors shaded; the three classical baselines are shown in italics to the right of the divider. \textbf{b} Final-round contributions without and with the punishment option, with the standard deviation across runs; $\ast$ marks a Holm-significant decrease under the exact test. \textbf{c} The share of runs in which punishment is purchased (bars) and the points spent per agent per round (points, right axis).}\label{fig:pgg}
\end{figure*}

Costly punishment inverts the institution it models (Fig.~\ref{fig:pgg}). In human experiments the option to punish \citep{pgg2024}
raises contributions; here final-round contributions fall in four of the nine models for which the
contrast is computable, by as much as 0.53 of endowment. The transcripts show why: punishment is
purchased in a median 97\% of runs, at a median of 2.5 points per agent per round, and the models
that spend most are the models whose contributions fall furthest (Spearman $\rho = -0.84$,
$p = 0.001$, $k = 11$). Two models, Qwen3-14B and Qwen3-8B, never purchase punishment in any run,
and they are exactly the two whose contributions do not move at all.


\input{tables/fidelity}

\subsection{Robustness: form matters, content matters more}\label{sec:rq2}
\begin{figure}[t]\centering
\includegraphics[width=\linewidth]{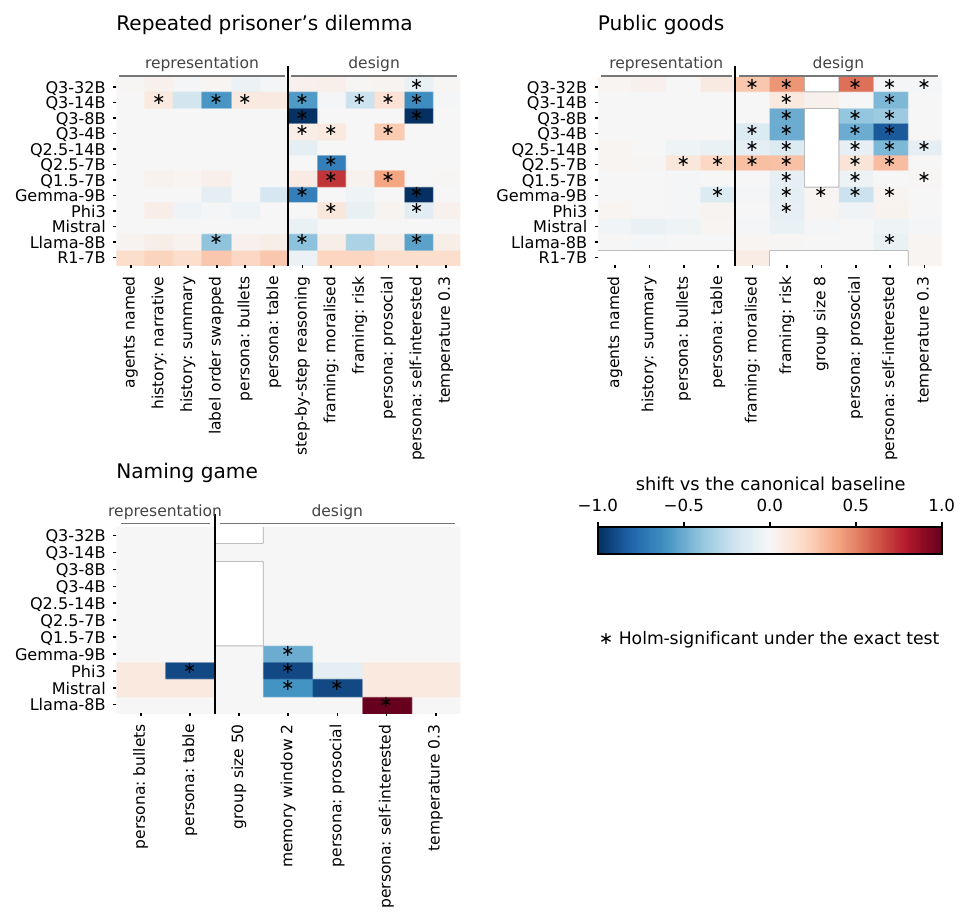}
\caption{Every cell of the audit grid. One panel per environment; rows are models, columns are perturbation cells, ordered representation-level first and separated from the design-level cells by a rule. Colour is the shift of the environment's primary outcome relative to the canonical baseline; $\ast$ marks a Holm-significant contrast under the exact test. The one-shot games, whose cells hold two runs and are descriptive.}\label{fig:heatmap}
\end{figure}

Figure~\ref{fig:heatmap} shows every cell of the audit grid and Table~\ref{tab:robustness} summarises
it. Across the three environments in which run-level tests are possible, at least one perturbation
moves behaviour significantly in 23 of 27 model-by-environment blocks; the one-shot games hold two
runs per cell and are reported descriptively throughout.

Separating the two kinds of perturbation changes the picture that a single fragility score would give.
Design-level cells --- those that alter what the agents are told --- move behaviour in 56 of 111
computable contrasts. Representation-level cells, which re-render information the agents already
have, move it in 8 of 71. Fragility is therefore not uniform: most of it is a response to content,
which is defensible, and a minority is a response to form, which is not.

That minority is nonetheless severe. Swapping the order in which the two actions are listed, leaving
every word and every payoff untouched, costs Qwen3-14B 58 points of cooperation
($\delta = -1.00$, $p_H = 2.8\times10^{-8}$) and Llama-3.1 40 points
($\delta = -0.68$, $p_H = 2.4\times10^{-3}$). Rendering the persona as a table rather than as a
sentence ends convention formation in Phi-3 outright, from 0.90 to 0.00
($\delta = -0.90$, $p_H = 4.7\times10^{-6}$), and raises Qwen2.5-7B's public-goods contributions by
21 points of endowment. Nothing in these manipulations carries information about the game.

Among the design-level cells, the persona content is the most destructive dimension. A
self-interested persona removes cooperation entirely in Gemma-2 and Qwen3-8B (both
$\delta = -1.00$, $p_H < 10^{-7}$) and, in the naming game, takes Llama-3.1 from never forming a
convention to always forming one ($\delta = +1.00$, $p_H = 2.4\times10^{-9}$). Asking a model to
reason step by step before answering is an intervention of the same magnitude: it costs Qwen3-8B 99
points of cooperation. A methodological choice that a researcher would describe as neutral --- how
the agent is asked to think --- is as consequential as the incentives.

Heterogeneity between models is total. $I^{2}$ reaches $100\%$ on four of the five primary outcomes
and $90.8\%$ on the fifth, the 11--20 game --- uninformative at these sample sizes, because with 30
runs per model and almost no within-model variance any difference between model means saturates it.
The useful summary is the spread of those means on the outcome scale. Cooperation rates at the canonical baseline span the entire range from
0.00 to 1.00 with a between-model standard deviation of 0.36; public-goods contributions span 0.02
to 1.00; ultimatum offers span 0.20 to 0.60. A result obtained with one model is not evidence about
language-model agents in general, and the practice of reporting a single model's society as such is
not defensible on this evidence.

 Fragility also cuts the other way: Llama-3.1's baseline \emph{failure} to form conventions is reversed to $100\%$ convergence by at least one perturbation cell --- null results are as prompt-sensitive as positive ones.

Statements about ``language-model agents'' as a class, extrapolated from any single model, therefore have close to no external validity in these settings.

\input{tables/robustness}

\begin{figure*}[t]\centering
\includegraphics[width=\linewidth]{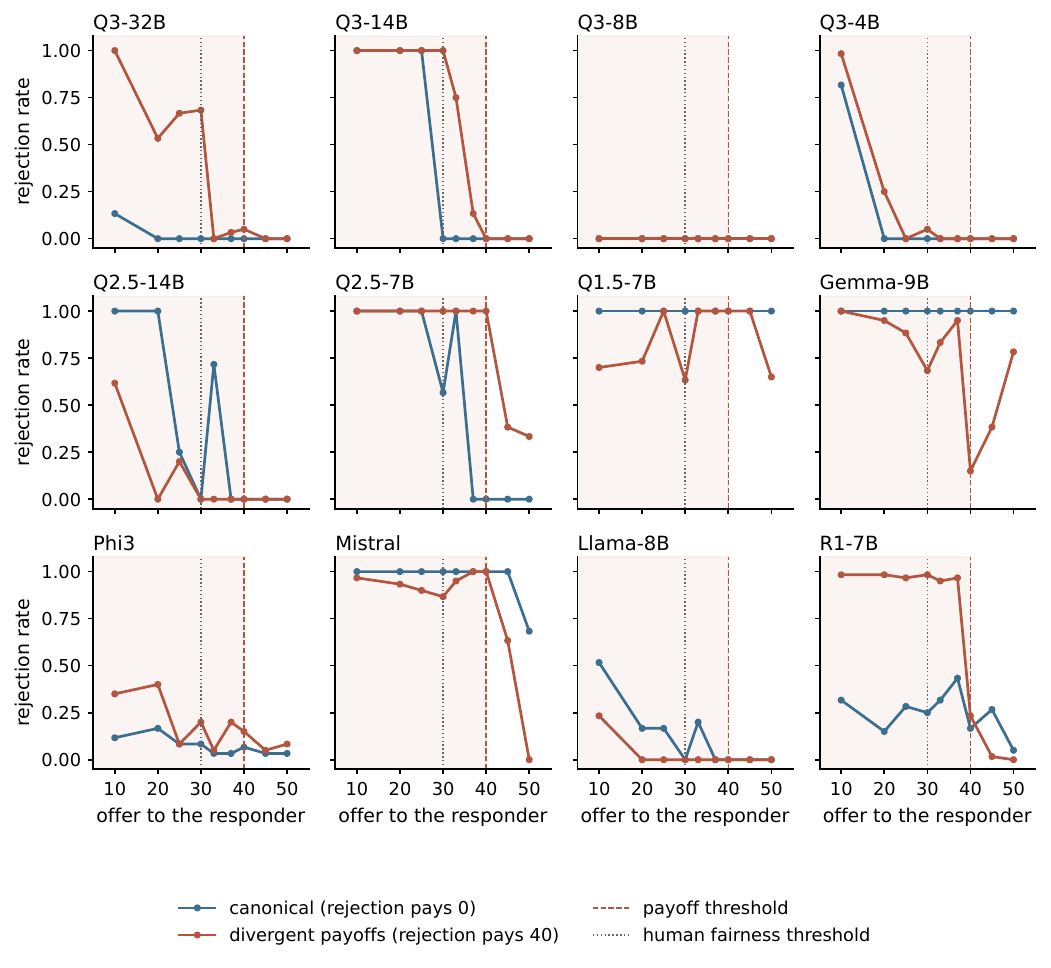}
\caption{Acceptance functions in the ultimatum game under a fixed schedule of offers. Each panel shows one model's rejection rate against the offer, in the canonical game where rejection pays nothing and in the divergent-payoff variant where rejection pays 40. The dashed line marks the offer above which accepting is the profitable act in the divergent variant; the dotted line marks the low-offer threshold conventionally reported for human responders. Shading marks the region in which rejection is payoff-maximising under the divergent payoffs.}\label{fig:offergrid}
\end{figure*}
\begin{figure*}[t]\centering
\includegraphics[width=\linewidth]{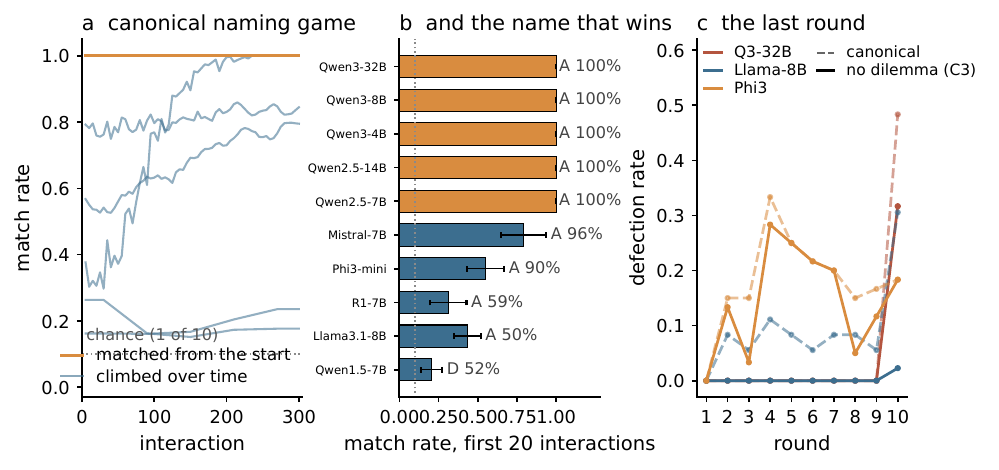}
\caption{How conventions form, and what survives when the incentive to defect is removed. \textbf{a} Match rate over the first 300 interactions of the canonical naming game, one line per model. \textbf{b} Match rate over the first twenty interactions with the name that wins. \textbf{c} Defection rate by round in the canonical game (dashed) and in the variant where cooperation is strictly dominant (solid).}\label{fig:residue}
\end{figure*}

\subsection{Provenance: retrieval, thresholds, and non-response}\label{sec:rq3}
The divergent-payoff ultimatum cell separates a memorised threshold from a computed one only for
offers between 30 and 40 points: below 30 both accounts reject, above 40 both accept. Because the
proposer is the same model as the responder, whether a model ever faces such an offer is itself an
outcome, and in the original design seven of nine models never did (Supplementary
Fig.~S2). Presenting responders with a
fixed schedule of offers removes that dependence. The resulting acceptance functions
(Fig.~\ref{fig:offergrid}) are given in full in Supplementary Table~S5, and they divide the roster
by how the acceptance threshold responds when rejection starts paying 40 points.

Only one model places that threshold where the outside option puts it. R1-Distill, the sole
reasoning-trained model, rejects offers of 10 to 37 in more than 95\% of games and accepts 40 and
above, giving a payoff consistency of 0.98 against a roster median of 0.61. Two more move the
threshold in the right direction without reaching it: Qwen3-14B from 25 to 33 and Qwen2.5-7B from 33
to 40. Two move it the wrong way, Qwen2.5-14B from 33 down to 10 and Mistral from 50 to 45, and
three do not move it at all. Qwen3-32B is the clearest case of a memorised boundary: it has no
threshold in the canonical game and acquires one at 30 in the divergent variant --- the human
low-offer threshold of roughly a third of the pie, not the 37 the incentive specifies --- so it
rejects 10 to 30 while accepting 33 and 37, forfeiting 7 and 3 points respectively.

Three models never produce a threshold in the divergent variant at all. Qwen3-8B accepts all 540
games in both payoff regimes, taking 10 points rather than a guaranteed 40 in every one of them;
Phi-3 and Llama-3.1 reject too little, too unsystematically, for any boundary to be located. At the
other extreme, Qwen1.5 and Gemma-2 reject essentially every offer in the canonical game, where
accepting always pays. An aggregate rejection rate cannot distinguish these cases; it reads the first as a script-follower
and the second as a payoff-follower.

The re-skinned variants C1 and C2, and what the models say when asked directly whether they recognise the paradigm, are reported in Supplementary Table~S3: recognition predicts none of the behavioural columns.

Convention formation tells a different story (Fig.~\ref{fig:residue}). Six models reach a convention at 50 interactions,
which is the earliest point the convergence rule can detect one; for the five whose trajectories
were retained, every one of the first twenty interactions is already a match, always on the first
name in the list, where chance is one in ten. Presenting each agent with its own permutation of the
same pool leaves coordination intact but slows it: those five now require 75 to 129 interactions,
and Gemma-2 rises from 130 to 250, while all of them still converge on the same name --- so the
shared prior is over the label rather than over list position. One model, Qwen2.5-7B, converges in only two of three seeds and takes 14 to 24
times longer. Even so, every model that converges still does so faster than the classical minimal naming
game \citep{baronchelli2006}, which reaches consensus in a median of 2{,}076 interactions: between
twenty-eight times faster for the quickest and twice as fast for Qwen2.5-7B, the slowest.

\begin{figure*}[t]\centering
\includegraphics[width=\textwidth]{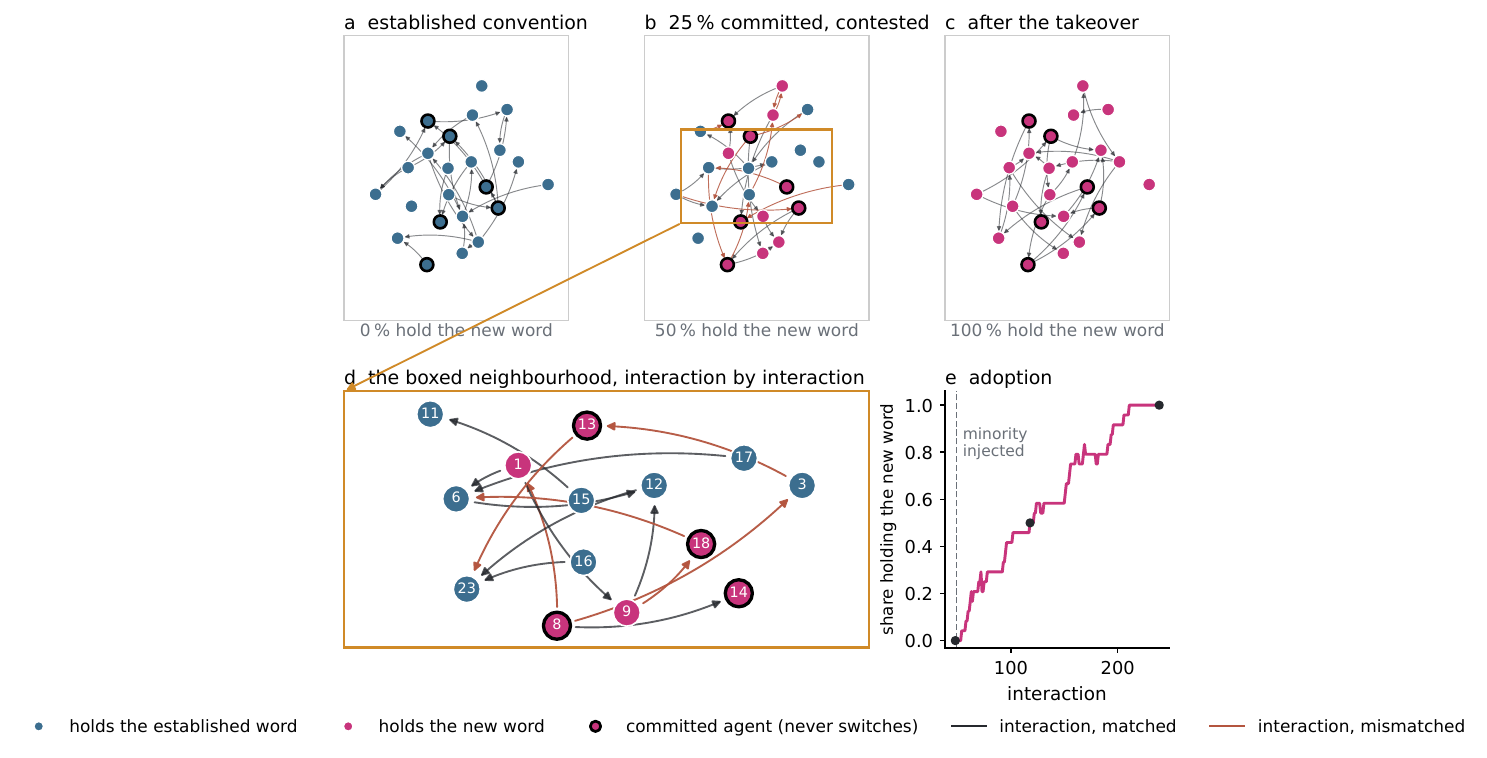}
\caption{One society overturning its own convention. The naming game is the only environment here
whose record is natively a graph: each line of the transcript is a directed interaction between two
identified agents, carrying the word each proposed and whether the two matched. \textbf{a--c} Three
moments of a single run of the tipping arm (Qwen3-32B, seed 0, 24 agents): the established
convention, the point at which a committed minority of six agents --- 25\% of the population, the
human critical mass --- has split the population evenly, and the state after the minority has
carried it. Node positions are computed once from the whole run and held fixed, so the only thing
that changes between panels is colour; edges show the last 25 interactions before each moment.
\textbf{d} The boxed neighbourhood at the contested moment, with individual interactions drawn.
\textbf{e} Adoption of the new word over the run. Committed agents, ringed in black, never switch;
mismatched interactions are the mechanism by which the rest do.}
\label{fig:society}
\end{figure*}

The tipping arm makes the same point in a form the other environments cannot: because the naming
game records who spoke to whom, a run can be read as a social process rather than as a curve.
Figure~\ref{fig:society} follows one. An established convention, held by all 24 agents, is contested
by six committed agents --- 25\% of the population, the human critical mass --- and is carried
completely 162 interactions later. The mechanism is visible in the failed interactions:
the committed agents never switch, and every mismatch they generate is an opportunity for an
uncommitted agent to change. It is worth being clear about what this does and does not show. It is
one run of one model, chosen because it converges; it is an illustration of the dynamics behind
Supplementary Fig.~S3, not additional evidence. The quantitative claim is coarser and less flattering: a critical mass could be estimated for seven of the twelve models, and none of the seven falls inside the human bracket of 21 to 25\%. Two tip at 20\% and five need 30\% or more; in two further models the convention is never overturned at any fraction tested, and in two others no convention forms to overturn (Supplementary Fig.~S3).

Where the memorised experiment does survive intact, it survives locally. In the variant where
cooperation is strictly dominant in every round, Qwen3-32B never defects in rounds one to nine and
defects in 32\% of final rounds --- an endgame effect in an environment with no endgame incentive.

\subsection{Scale, generation and reasoning}\label{sec:axes}
\begin{figure*}[t]\centering
\includegraphics[width=\textwidth]{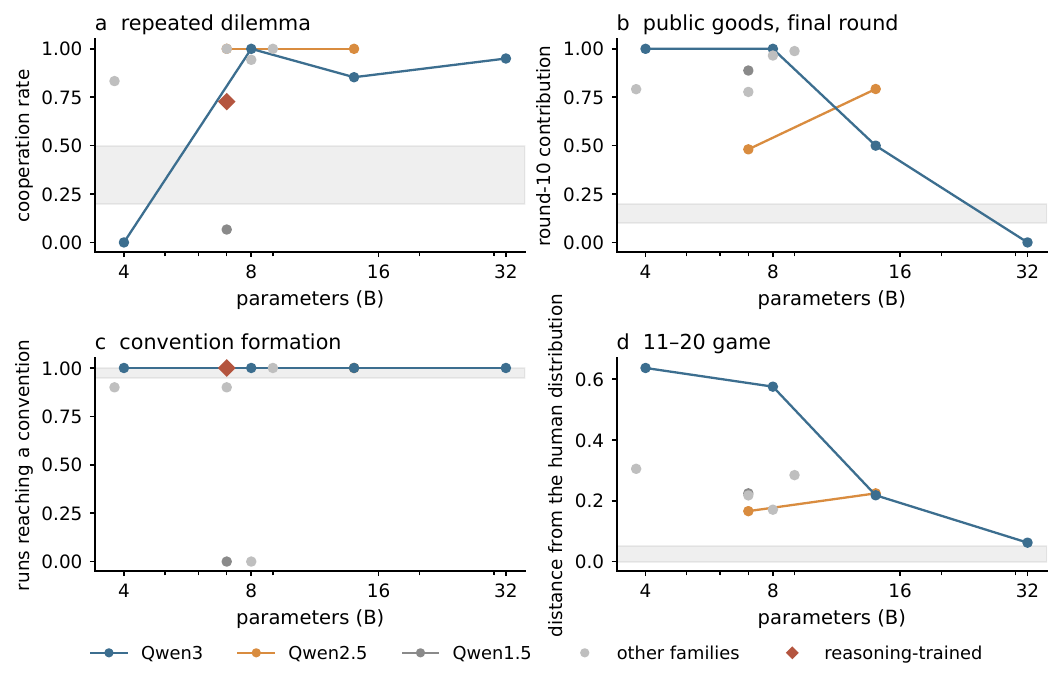}
\caption{Four primary outcomes against model size. \textbf{a} Cooperation rate in the repeated
dilemma. \textbf{b} Final-round contribution in the public-goods game. \textbf{c} Share of runs
reaching a convention. \textbf{d} Distance between the model's choice distribution in the 11--20
game and the human one. Lines join models of the same family; the reasoning-trained model is drawn
as a diamond; shaded bands mark the human values. Public-goods cells holding fewer than three runs
are omitted.}
\label{fig:scale}
\end{figure*}

Within the Qwen3 family, strategic depth improves monotonically with size (Fig.~\ref{fig:scale}): the distance between the
model's choice distribution in the 11--20 game and the human one falls from 0.637 at 4B through
0.576 and 0.218 to 0.062 at 32B, the closest approach to a human distribution anywhere in the study.
No comparable ordering appears elsewhere. On public-goods contributions the same family moves in the
opposite direction --- final-round contributions fall from the full endowment at 4B and 8B to 0.50 at
14B and 0.00 at 32B --- while the Qwen2.5 family moves the other way, from 0.48 at 7B to 0.79 at
14B. Larger models are closer to humans on one outcome, further on another, and the direction is not
even consistent across families from the same laboratory.

Generation is equally unstable. Three releases from one laboratory at comparable size invert baseline
outcomes rather than refining them: Qwen1.5-7B cooperates in 0.07 of prisoner's dilemma rounds and
Qwen2.5-7B in 1.00 of them.

The reasoning axis rests on a single model and is reported as exploratory. It is nonetheless the
sharpest result in the contamination audit: R1-Distill is the only model in the roster whose
acceptance threshold sits where the outside option puts it, and the only one whose time to a
convention is unchanged when the name pool is shuffled. Even there the independence is partial:
its speed does not depend on list position, but the name it settles on does, moving from A under
a fixed order to E under a shuffled one. What reasoning training appears to buy is not similarity
to humans --- R1 is no closer to the human corridor than its
neighbours --- but responsiveness to the situation actually presented.

%% file: tables/glance.tex
\begin{table*}[t]
\caption{The study at a glance. Each row states one research question, the evidence assembled for it, and the highest certification tier the evidence supports. Denominators are the number of models for which the comparison is computable; R1-Distill completed a partial battery and is excluded from cells where it holds fewer than three runs. Equivalence is the two one-sided test against the human anchor at the margin fixed in advance; significance is Holm-adjusted within family under the exact test. Per-model acceptance thresholds and payoff consistency are in Supplementary Table~S5; a high consistency score can be produced by rejecting almost everything, so the threshold rather than the score is the quantity reported here.}
\label{tab:glance}\footnotesize\centering
\begin{tabular}{@{}p{1.85cm} p{2.45cm} p{6.75cm} l@{}}
\toprule
Question & What it asks & What the data show & Tier \\
\midrule
RQ1 Level fidelity & Does behaviour match human levels? & public goods round 1: 8/11 equivalent; round 10: 0/11; ultimatum offers: 1/12; dictator: 0/12; PD inside the human corridor: 0/12 & Tier 1 \\[2pt]
RQ1 Directional fidelity & Does behaviour move in the human direction? & public goods decay 1/12; PD decay 3/12; convention formation 10/12 & Tier 1 \\[2pt]
RQ2 Robustness & Does the finding survive re-rendering the same rules? & Holm-significant shifts in 64 of 182 computable cells; largest representation-level reversal 1.00; largest design-level 1.00 & Tier 1 \\[2pt]
RQ3 Contamination & Is the behaviour retrieved or computed? & ultimatum, under a fixed offer schedule: 1 of 12 models sets its acceptance threshold where the outside option requires, 2 move theirs partway, 2 the wrong way and 3 never acquire one; naming with a per-agent shuffled pool: 7 of 8 models still converge in every seed, 1.5 to 2.6 times slower, while Qwen2.5-7B converges in only 2 of 3 seeds at 14--24 times the fixed-order value; final-round defection with no endgame incentive in 3 of 10 models & Tier 1 \\[2pt]
RQ4 Certification & What can the instrument certify? & no finding reaches Tier 3. Convention formation reaches Tier 2: the shuffled-pool control shows that coordination is driven by a shared prior over the labels rather than by shared list position, and that negotiation does occur once that cue is removed --- but still between twice and twenty-eight times faster than the classical baseline, so the mechanism is related to the human one without matching it. Every other finding is Tier 1 & -- \\[2pt]
\bottomrule
\end{tabular}
\end{table*}

%% file: tables/fidelity.tex
\begin{table*}[t]
\caption{Baseline fidelity against the human anchors, one row per model. Values are means over the canonical baseline runs with the standard deviation across runs; $n$ is the number of runs. \textbf{Bold} marks an outcome that passes the two one-sided equivalence test against its anchor at the margin fixed in advance. PD cooperation is compared with the corridor $0.20$--$0.50$ reported across prisoner's dilemma studies; public-goods contributions with $0.50$ in round~1 and $0.15$ in round~10; ultimatum offers with $0.40$ of the pie; convention formation with the human result that a convention always forms; and the 11--20 choice distribution through the Wasserstein distance to the distribution of Arad and Rubinstein, normalised by the width of the support, with a bootstrap 95\% interval. R1-Distill completed a partial battery and its public-goods cell holds a single run, shown as --. In the $p^{*}$ column, a value is the smallest tested committed-minority fraction at which the convention is overturned in at least half of runs; $>0.40$ means a convention forms but was never overturned at any tested fraction; ``no conv.'' means no convention forms, so there is nothing to overturn; and ``not run'' means the tipping arm was not executed for that model.}
\label{tab:fidelity}\scriptsize\setlength{\tabcolsep}{2.6pt}\centering
\resizebox{\textwidth}{!}{%
\begin{tabular}{l cc cc c cc c}
\toprule
 & \multicolumn{2}{c}{Repeated dilemma} & \multicolumn{2}{c}{Public goods} & Ultimatum & \multicolumn{2}{c}{Naming} & 11--20 \\
\cmidrule(lr){2-3}\cmidrule(lr){4-5}\cmidrule(lr){6-6}\cmidrule(lr){7-8}\cmidrule(lr){9-9}
Model & cooperation & in band & round 1 & round 10 & offer & conventions & $p^{*}$ & $W_1$ \\
\midrule
Qwen3-32B & 0.95\,$\pm$\,0.05 & no & 0.15\,$\pm$\,0.17 & 0.00\,$\pm$\,0.00 & 0.49\,$\pm$\,0.01 & 1.00 & 0.20 & 0.062 [0.06,0.07] \\
Qwen3-14B & 0.85\,$\pm$\,0.08 & no & \textbf{0.50\,$\pm$\,0.00} & 0.50\,$\pm$\,0.00 & 0.50\,$\pm$\,0.00 & 1.00 & 0.40 & 0.218 [0.21,0.22] \\
Qwen3-8B & 1.00\,$\pm$\,0.00 & no & 1.00\,$\pm$\,0.00 & 1.00\,$\pm$\,0.00 & 0.22\,$\pm$\,0.05 & 1.00 & 0.30 & 0.576 [0.58,0.58] \\
Qwen3-4B & 0.00\,$\pm$\,0.00 & no & 1.00\,$\pm$\,0.00 & 1.00\,$\pm$\,0.00 & 0.50\,$\pm$\,0.00 & 1.00 & 0.30 & 0.637 [0.63,0.64] \\
Qwen2.5-14B & 1.00\,$\pm$\,0.00 & no & \textbf{0.57\,$\pm$\,0.02} & 0.79\,$\pm$\,0.09 & \textbf{0.40\,$\pm$\,0.00} & 1.00 & 0.20 & 0.224 [0.22,0.22] \\
Qwen2.5-7B & 1.00\,$\pm$\,0.00 & no & \textbf{0.40\,$\pm$\,0.00} & 0.48\,$\pm$\,0.04 & 0.60\,$\pm$\,0.00 & 1.00 & 0.40 & 0.166 [0.16,0.18] \\
Qwen1.5-7B & 0.07\,$\pm$\,0.06 & no & \textbf{0.50\,$\pm$\,0.00} & 0.89\,$\pm$\,0.03 & 0.50\,$\pm$\,0.00 & 0.00 & no conv. & 0.224 [0.22,0.22] \\
Gemma2-9B & 1.00\,$\pm$\,0.00 & no & \textbf{0.50\,$\pm$\,0.04} & 0.99\,$\pm$\,0.02 & 0.22\,$\pm$\,0.01 & 1.00 & > 0.40 & 0.284 [0.26,0.31] \\
Phi3-mini & 0.83\,$\pm$\,0.08 & no & \textbf{0.53\,$\pm$\,0.05} & 0.79\,$\pm$\,0.07 & 0.49\,$\pm$\,0.01 & 0.90 & 0.40 & 0.305 [0.27,0.34] \\
Mistral-7B & 1.00\,$\pm$\,0.00 & no & \textbf{0.54\,$\pm$\,0.08} & 0.78\,$\pm$\,0.10 & 0.51\,$\pm$\,0.01 & 0.90 & > 0.40 & 0.218 [0.22,0.22] \\
Llama3.1-8B & 0.94\,$\pm$\,0.19 & no & \textbf{0.55\,$\pm$\,0.10} & 0.97\,$\pm$\,0.04 & 0.47\,$\pm$\,0.02 & 0.00 & no conv. & 0.171 [0.16,0.18] \\
R1-7B & 0.73\,$\pm$\,0.30 & no & -- & -- & 0.20\,$\pm$\,0.05 & 1.00 & not run & -- \\
\bottomrule
\end{tabular}}
\end{table*}

%% file: tables/robustness.tex
\begin{table*}[t]
\caption{Robustness of the primary outcome under the perturbation library, one row per model, split by the level of the perturbation. $\Delta_{\max}$ (Eq.~\ref{eq:swing}) is the range of cell means over the canonical baseline and either the representation-level cells (the same rules re-rendered: persona layout, history rendering and summary, action-label order, agent naming) or the design-level cells (content changes: persona content, moralized or risk framing, step-by-step reasoning, sampling temperature, memory window, group size). $S$ is the share of cells (baseline included) in which the canonical human effect keeps its sign (PD and PGG: decline of cooperation or contributions over rounds; naming: convergence); it is defined only for those three environments. Perturbation cells hold 10 runs per model (2 runs of 50 games in the one-shot games, where $\Delta_{\max}$ is therefore descriptive); baseline cells 30 (6 in the one-shot games). -- $=$ arm not run, baseline with fewer than 3 runs, or fewer than two cells with at least 2 runs (R1-Distill completed a partial battery; its per-cell run counts are listed in the Supplementary roster).}
\label{tab:robustness}\scriptsize\setlength{\tabcolsep}{2.2pt}\centering
\resizebox{\textwidth}{!}{%
\begin{tabular}{l ccccc ccccc ccc}
\toprule
 & \multicolumn{5}{c}{$\Delta_{\max}$, representation-level} & \multicolumn{5}{c}{$\Delta_{\max}$, design-level} & \multicolumn{3}{c}{$S$, representation\,/\,design} \\
\cmidrule(lr){2-6}\cmidrule(lr){7-11}\cmidrule(lr){12-14}
Model & PD & PGG & Ult. & 11--20 & Naming & PD & PGG & Ult. & 11--20 & Naming & PD & PGG & Naming \\
\midrule
Qwen3-14B & 0.68 & 0.00 & 0.00 & 0.08 & 0.00 & 0.77 & 0.55 & 0.00 & 0.01 & 0.00 & 0.57\,/\,0.57 & 0.00\,/\,0.00 & 1.00\,/\,1.00 \\
Gemma2-9B & 0.17 & 0.14 & 0.06 & 0.02 & 0.00 & 1.00 & 0.24 & 0.20 & 0.01 & 0.50 & 0.00\,/\,0.00 & 0.00\,/\,0.00 & 1.00\,/\,1.00 \\
Llama3.1-8B & 0.46 & 0.02 & 0.03 & 0.08 & 0.00 & 0.59 & 0.08 & 0.12 & 0.08 & 1.00 & 0.14\,/\,0.43 & 0.00\,/\,0.00 & 0.00\,/\,0.17 \\
Mistral-7B & 0.00 & 0.09 & 0.21 & 0.00 & 0.10 & 0.07 & 0.07 & 0.04 & 0.00 & 1.00 & 0.00\,/\,0.00 & 0.00\,/\,0.00 & 1.00\,/\,0.67 \\
Phi3-mini & 0.11 & 0.04 & 0.02 & 0.03 & 1.00 & 0.23 & 0.12 & 0.02 & 0.02 & 1.00 & 0.57\,/\,0.57 & 0.00\,/\,0.00 & 0.67\,/\,0.83 \\
Qwen3-32B & 0.10 & 0.10 & 0.01 & 0.07 & 0.00 & 0.12 & 0.57 & 0.17 & 0.44 & 0.00 & 0.57\,/\,0.29 & 1.00\,/\,0.50 & 1.00\,/\,1.00 \\
Qwen3-8B & 0.00 & 0.00 & 0.48 & 0.00 & 0.00 & 1.00 & 0.50 & 0.57 & 0.00 & 0.00 & 0.00\,/\,0.00 & 0.00\,/\,0.33 & 1.00\,/\,1.00 \\
Qwen3-4B & 0.00 & 0.00 & 0.49 & 0.00 & 0.00 & 0.26 & 0.84 & 0.49 & 0.00 & 0.00 & 0.00\,/\,0.29 & 0.00\,/\,0.17 & 1.00\,/\,1.00 \\
Qwen2.5-14B & 0.00 & 0.08 & 0.00 & 0.00 & 0.00 & 0.09 & 0.45 & 0.17 & 0.00 & 0.00 & 0.00\,/\,0.00 & 0.00\,/\,0.17 & 1.00\,/\,1.00 \\
Qwen2.5-7B & 0.00 & 0.21 & 0.00 & 0.00 & 0.00 & 0.68 & 0.31 & 0.20 & 0.00 & 0.00 & 0.00\,/\,0.14 & 0.00\,/\,0.00 & 1.00\,/\,1.00 \\
Qwen1.5-7B & 0.06 & 0.02 & 0.00 & 0.00 & 0.00 & 0.71 & 0.10 & 0.00 & 0.00 & 0.00 & 1.00\,/\,1.00 & 0.00\,/\,0.00 & 0.00\,/\,0.00 \\
R1-7B & 0.27 & -- & 0.00 & -- & -- & 0.27 & -- & 0.02 & -- & -- & 0.00\,/\,0.00 & --\,/\,-- & --\,/\,-- \\
\bottomrule
\end{tabular}}
\end{table*}

%% file: sections/intro_related_discussion_part2.tex
\section{Discussion}\label{sec:discussion}

\subsection{What a social scientist can and cannot do with these societies}
The results support a narrow and specific use. Populations of language-model agents reproduce the
\emph{starting points} of several classic experiments: first-round contributions to a public good
fall inside the equivalence margin for eight of eleven models, and conventions form in ten of
twelve. They do not reproduce what happens next. Contributions do not decay, cooperation does not
decline, and the institution of costly punishment --- which raises contributions in human groups ---
lowers them here in four of the nine models where the comparison can be made. A researcher who uses
these societies to generate a hypothesis about how an interaction begins is on defensible ground. A
researcher who uses them to predict how an interaction develops, or to estimate the effect of an
institution, is not.

The reason is visible in the transcripts rather than in the outcome statistics. In the public-goods
game the models that punish most are the models whose contributions collapse furthest
($\rho = -0.84$), and the two models whose contributions do not move are exactly the two that never
purchase punishment at all. The mechanism is present but miscalibrated: the agents understand that
punishment is available and use it, and the group response to being punished is the opposite of the
human one. This is a more useful finding for a social scientist than a simple mismatch of levels,
because it says which part of the model of human behaviour is missing.

\subsection{Thresholds, and what they are made of}
Presenting responders with a fixed schedule of offers separates behaviours that any aggregate
statistic merges. One model of twelve sets its rejection threshold where the outside option puts it.
Two move theirs in the direction the incentive requires without reaching it, two move theirs the
wrong way, and three never acquire a threshold at all --- one of those accepting every offer in both
payoff regimes, taking 10 points rather than a guaranteed 40 in every one of 540 games. At the other
extreme two models reject essentially every offer in the canonical game, where accepting always
pays. The most instructive case is the model whose threshold appears only when the payoffs change
and then lands on the human value rather than the one the incentive specifies.

Only a threshold that both exists and moves with the incentive is evidence of anything like a
preference over outcomes, and one model of twelve has one. A threshold that appears at the human
value regardless of what rejection pays is a memorised boundary. A model with no threshold at all is
not responding to the situation, and an aggregate rejection rate reads it as either a
script-follower or a payoff-follower depending on which way it fails. The
practical implication is that a study which reports mean behaviour in a bargaining game with
language-model agents cannot distinguish a preference from a non-response, and should not be read as
if it could.

\subsection{An instrument can produce a confident and wrong mechanism claim}
The most instructive result of this work is one the instrument produced against itself. Six models
form a naming convention at the earliest interaction the convergence rule can detect, and for the
five whose trajectories were retained, every one of the first twenty interactions already matches.
Read from the outcome statistics alone, that is a striking finding: coordination without negotiation,
forty times faster than the classical minimal naming game. The natural reading, and the one this
study first drew, is that convention formation in these societies is retrieval rather than
emergence.

The agents were, however, all shown the same ten names in the same order, and the fallback for an
unparsable reply was the first name in the list. Instant agreement on the first-listed item is
equally consistent with a shared prior about the task and a shared prior about the list. Presenting
each agent with its own permutation of the same pool separates them, and the answer is neither of
the simple ones: coordination survives --- the models still converge, and still on the same name, so
the prior is over the label --- but it now takes 75 to 129 interactions instead of the 50 that the
rule could first detect, and one model converges in only two of three runs at 14 to 24 times the
original figure. The negotiation was real; it had been hidden by a cue that no one had thought to
vary.

This has a general moral for benchmark design that goes beyond the present study. A benchmark that
measures outcomes can support a mechanism claim only if it also varies the features of its own
apparatus that could produce the same outcome for a different reason. The perturbation library in
this instrument was built to test whether findings survive changes to the apparatus; it did not
contain the one manipulation that would have caught this, and the manipulation was added only after
a transcript audit made the alternative explanation visible. Instruments of this kind should be
expected to include such controls, and their absence should be treated as a limitation on what can
be concluded, not as a detail of implementation.

\subsection{What a Tier-3 result would require}
No finding in this study reaches the top of the certification ladder, and it is worth stating what
would. A transferable claim requires three things at once: equivalence with the human anchor rather
than agreement in direction; survival of the perturbation library, including the representation-level
cells that carry no information; and a transcript audit consistent with the mechanism recorded in
the human experiment. Convention formation is the only finding that reaches Tier~2, and it is worth showing why rather than asserting it: sign stability is at or above $0.8$ at both perturbation levels in seven models spanning three families --- Qwen3, Qwen2.5 and Gemma-2, exactly the minimum the criterion requires --- and the finding is not reversed when the payoffs reward mismatching. It comes closest of any result to the third condition, now that the shuffled-pool control shows genuine negotiation, but the negotiation still runs between twice and twenty-eight times faster than the classical dynamics, so the mechanism is related to the human one without matching it. No other finding satisfies more than one condition.

\subsection{Limits of this evidence}
The anchors are summary-level, and they describe particular people. Equivalence against a meta-analytic mean is a weaker test than equivalence against a distribution of individual human decisions, and three of the seven --- the 11--20 choice distribution, the tipping threshold, and the public-goods punishment value --- rest on a single experiment rather than on a synthesis. The underlying experiments were also run overwhelmingly on Western, educated, industrialised, rich and democratic samples --- university participants in Europe, North America and Israel --- so ``human'' here means those populations. Where cross-cultural work exists it finds substantial variation in exactly the quantities used as anchors, most clearly in ultimatum offers and rejection rates. A model that matched these anchors perfectly would be matching a narrow slice of humanity, and a model that misses them is not thereby shown to be unlike people in general. One anchor was found to have been mis-transcribed
in two cells during this work, which is itself an argument for releasing anchors as data with
provenance rather than as constants in a paper.

The roster is open-weight models small enough to run on one consumer graphics card. That choice buys
exact reproducibility and independence from provider APIs, and it costs generality: nothing here
establishes how frontier models behave, and the one reasoning-trained model in the roster is a
distillation rather than a frontier reasoning system. The reasoning axis therefore rests on a single
model and is reported as exploratory throughout.

The one-shot battery holds two runs per perturbation cell, which supports description and not
inference, and it is labelled as such wherever it appears. The public-goods punishment anchor comes
from a single experiment. Finally, this is single-author work: the analysis code was reviewed
against the released run record rather than by an independent analyst, and the corrections
documented here were found by auditing transcripts against outcome statistics, a procedure this
article recommends precisely because it worked.

\section{Conclusion}\label{sec:conclusion}
Silicon societies currently reproduce where human interactions begin and not how they proceed, they
respond to the form of the apparatus as well as to its content, and in the situations where the
memorised experiment and the payoffs disagree most of them do not produce a decision function over
the situation at all. On the certification ladder defined here, that supports exploratory claims and
no more --- which is a useful conclusion rather than a discouraging one, because the instrument also
shows which specific component of human social behaviour each model is missing, and because the one
manipulation that overturned this study's own mechanism claim cost a few hours of computation. The
benchmark, the environments, the anchors with their provenance, and the complete run record are
released so that the next instrument of this kind can begin where this one ends.

%% file: tables/si_roster.tex
\begin{table*}[htbp]
\caption{Model roster and execution statistics. Parameter counts are the published totals. Quantisation is as served; the Qwen3 models expose an optional deliberation mode, which was disabled throughout, so the reasoning axis rests on the single reasoning-trained model rather than on a decoding switch. Attempted calls are estimated from the run manifest; invalid actions are replies that failed schema validation and were replaced by the environment's fallback. R1-Distill completed a partial battery, which is why its cell count is lower.}
\label{tab:si-roster}\scriptsize\setlength{\tabcolsep}{3pt}\centering
\resizebox{\textwidth}{!}{%
\begin{tabular}{l r l l r r r r r r}
\toprule
Model & Params (B) & Quantisation & Deliberation & Runs & Cells & Calls & Invalid & Rate & GPU h \\
\midrule
Qwen3-32B & 32.8 & AWQ 4-bit & off & 792 & 86 & 1,699,840 & 10 & 0.00\% & 49.6 \\
Qwen3-14B & 14.8 & FP8 & off & 827 & 88 & 1,940,640 & 0 & 0.00\% & 131.8 \\
Qwen3-8B & 8.2 & bf16 & off & 792 & 86 & 1,699,840 & 0 & 0.00\% & 72.4 \\
Qwen3-4B & 4.0 & bf16 & off & 792 & 86 & 1,699,840 & 21 & 0.00\% & 47.3 \\
Qwen2.5-14B & 14.7 & AWQ 4-bit & off & 792 & 86 & 1,699,840 & 0 & 0.00\% & 16.3 \\
Qwen2.5-7B & 7.6 & AWQ 4-bit & off & 792 & 86 & 1,699,840 & 0 & 0.00\% & 21.6 \\
Qwen1.5-7B & 7.7 & bf16 & off & 792 & 86 & 1,699,840 & 765 & 0.05\% & 262.8 \\
Gemma2-9B & 9.2 & bf16 & off & 827 & 88 & 1,940,640 & 6 & 0.00\% & 113.7 \\
Phi3-mini & 3.8 & bf16 & off & 827 & 88 & 1,940,640 & 1,329 & 0.07\% & 109.1 \\
Mistral-7B & 7.2 & bf16 & off & 827 & 88 & 1,940,640 & 167 & 0.01\% & 106.3 \\
Llama3.1-8B & 8.0 & bf16 & off & 827 & 88 & 1,940,640 & 22 & 0.00\% & 254.5 \\
R1-7B & 7.6 & bf16 & off & 224 & 44 & 364,060 & 1,812 & 0.50\% & 301.6 \\
\bottomrule
\end{tabular}}
\end{table*}

%% file: tables/si_stats.tex
\begin{table*}[htbp]
\caption{Test-level record for the contrasts named in the main text, plus every contamination contrast: 95 of 227 computable contrasts, ordered by family. $p$ is the exact permutation value conditional on the observed ties and $p_H$ its Holm-adjusted counterpart within the family. The complete file, including the value each test would take under a normal approximation, is \texttt{results/stat\_tests.csv} in the released record.}
\label{tab:si-stats}\tiny\setlength{\tabcolsep}{3pt}\centering
\resizebox{\textwidth}{!}{%
\begin{tabular}{l l l l r r r r}
\toprule
Family & Env & Model & Contrast & $n$ & $\delta$ & $p$ & $p_H$ \\
\midrule
contamination & bargain & Gemma2-9B & C3 vs C0 & 6+6 & -1.00 & 0.002 & 0.015 \\
contamination & bargain & Mistral-7B & C3 vs C0 & 6+6 & -1.00 & 0.002 & 0.015 \\
contamination & bargain & Qwen1.5-7B & C3 vs C0 & 6+6 & -1.00 & 0.002 & 0.015 \\
contamination & bargain & Qwen2.5-7B & C3 vs C0 & 6+6 & +1.00 & 0.002 & 0.015 \\
contamination & bargain & Qwen3-32B & C3 vs C0 & 6+6 & +1.00 & 0.002 & 0.015 \\
contamination & bargain & Phi3-mini & C3 vs C0 & 6+6 & +0.67 & 0.067 & 0.134 \\
contamination & bargain & Llama3.1-8B & C3 vs C0 & 6+6 & -0.11 & 0.810 & 0.810 \\
contamination & naming & Gemma2-9B & C3 vs C0 & 30+30 & -1.00 & 1.7e-17 & 2.0e-16 \\
contamination & naming & Mistral-7B & C3 vs C0 & 30+30 & -1.00 & 1.7e-17 & 2.0e-16 \\
contamination & naming & Phi3-mini & C3 vs C0 & 30+30 & -1.00 & 1.7e-17 & 2.0e-16 \\
contamination & naming & Qwen2.5-14B & C3 vs C0 & 30+30 & -1.00 & 1.7e-17 & 2.0e-16 \\
contamination & naming & Qwen2.5-7B & C3 vs C0 & 30+30 & -1.00 & 1.7e-17 & 2.0e-16 \\
contamination & naming & Qwen3-14B & C3 vs C0 & 30+30 & -1.00 & 1.7e-17 & 2.0e-16 \\
contamination & naming & Qwen3-32B & C3 vs C0 & 30+30 & -1.00 & 1.7e-17 & 2.0e-16 \\
contamination & naming & Qwen3-4B & C3 vs C0 & 30+30 & -1.00 & 1.7e-17 & 2.0e-16 \\
contamination & naming & Qwen3-8B & C3 vs C0 & 30+30 & -1.00 & 1.7e-17 & 2.0e-16 \\
contamination & naming & R1-7B & C3 vs C0 & 30+30 & -1.00 & 1.7e-17 & 2.0e-16 \\
contamination & naming & Llama3.1-8B & C3 vs C0 & 30+30 & -0.85 & 1.1e-10 & 2.2e-10 \\
contamination & naming & Qwen1.5-7B & C3 vs C0 & 30+30 & -0.66 & 2.6e-06 & 2.6e-06 \\
contamination & pd & Qwen3-4B & C3 vs C0 & 30+30 & +1.00 & 1.7e-17 & 1.2e-16 \\
contamination & pd & Qwen3-14B & C3 vs C0 & 30+30 & +0.72 & 6.0e-08 & 3.6e-07 \\
contamination & pd & Qwen1.5-7B & C3 vs C0 & 30+30 & +0.51 & 1.4e-04 & 7.0e-04 \\
contamination & pd & R1-7B & C3 vs C0 & 28+11 & +0.68 & 2.6e-04 & 0.001 \\
contamination & pd & Llama3.1-8B & C3 vs C0 & 30+30 & +0.14 & 0.138 & 0.413 \\
contamination & pd & Phi3-mini & C3 vs C0 & 30+30 & +0.12 & 0.392 & 0.590 \\
contamination & pd & Qwen3-32B & C3 vs C0 & 30+30 & +0.17 & 0.295 & 0.590 \\
perturbation & naming & Llama3.1-8B & persona\_content=self\_interested vs b & 10+30 & +1.00 & 2.4e-09 & 2.4e-09 \\
perturbation & naming & Mistral-7B & persona\_content=prosocial vs baseline & 10+30 & -0.90 & 6.7e-07 & 4.7e-06 \\
perturbation & naming & Phi3-mini & memory=2 vs baseline & 10+30 & -0.90 & 6.7e-07 & 4.7e-06 \\
perturbation & naming & Phi3-mini & persona\_format=table vs baseline & 10+30 & -0.90 & 6.7e-07 & 4.7e-06 \\
perturbation & naming & Gemma2-9B & memory=2 vs baseline & 10+30 & -0.50 & 7.7e-04 & 7.7e-04 \\
perturbation & naming & Mistral-7B & memory=2 vs baseline & 10+30 & -0.60 & 0.001 & 0.007 \\
perturbation & pd & Qwen2.5-7B & framing=moralized vs baseline & 10+30 & -1.00 & 2.4e-09 & 2.4e-09 \\
perturbation & pd & Qwen3-8B & cot vs baseline & 10+30 & -1.00 & 2.4e-09 & 4.7e-09 \\
perturbation & pd & Qwen3-8B & persona\_content=self\_interested vs b & 10+30 & -1.00 & 2.4e-09 & 4.7e-09 \\
perturbation & pd & Qwen3-4B & framing=moralized vs baseline & 10+30 & +1.00 & 2.4e-09 & 7.1e-09 \\
perturbation & pd & Qwen3-4B & persona\_content=prosocial vs baseline & 10+30 & +1.00 & 2.4e-09 & 7.1e-09 \\
perturbation & pd & Gemma2-9B & persona\_content=self\_interested vs b & 10+30 & -1.00 & 2.4e-09 & 1.2e-08 \\
perturbation & pd & Qwen1.5-7B & framing=moralized vs baseline & 10+30 & +1.00 & 2.4e-09 & 2.8e-08 \\
perturbation & pd & Qwen3-14B & cot vs baseline & 10+30 & -1.00 & 2.4e-09 & 2.8e-08 \\
perturbation & pd & Qwen3-14B & label\_order=swapped vs baseline & 10+30 & -1.00 & 2.4e-09 & 2.8e-08 \\
perturbation & pd & Qwen3-14B & persona\_content=self\_interested vs b & 10+30 & -1.00 & 2.4e-09 & 2.8e-08 \\
perturbation & pd & Qwen1.5-7B & persona\_content=prosocial vs baseline & 10+30 & +0.99 & 7.1e-09 & 7.8e-08 \\
perturbation & pd & Qwen3-14B & framing=risk vs baseline & 10+30 & -0.94 & 4.7e-08 & 4.2e-07 \\
perturbation & pd & Llama3.1-8B & cot vs baseline & 10+30 & -0.88 & 1.4e-06 & 1.6e-05 \\
perturbation & pd & Gemma2-9B & cot vs baseline & 10+30 & -0.70 & 1.3e-05 & 5.1e-05 \\
perturbation & pd & Qwen3-14B & persona\_content=prosocial vs baseline & 10+30 & +0.83 & 7.1e-06 & 5.7e-05 \\
perturbation & pd & Llama3.1-8B & persona\_content=self\_interested vs b & 10+30 & -0.81 & 1.2e-05 & 1.4e-04 \\
perturbation & pd & Qwen3-4B & cot vs baseline & 10+30 & +0.50 & 7.7e-04 & 7.7e-04 \\
perturbation & pd & Llama3.1-8B & label\_order=swapped vs baseline & 10+30 & -0.68 & 2.4e-04 & 0.002 \\
perturbation & pd & Phi3-mini & framing=moralized vs baseline & 10+30 & +0.70 & 3.6e-04 & 0.004 \\
perturbation & pd & Phi3-mini & persona\_content=self\_interested vs b & 10+30 & -0.69 & 5.2e-04 & 0.006 \\
perturbation & pd & Qwen3-14B & history\_format=narrative vs baseline & 10+30 & +0.65 & 0.001 & 0.008 \\
perturbation & pd & Qwen3-14B & persona\_format=bullets vs baseline & 10+30 & +0.56 & 0.006 & 0.035 \\
perturbation & pd & Qwen3-32B & persona\_content=self\_interested vs b & 10+30 & -0.55 & 0.004 & 0.045 \\
perturbation & pgg & Qwen3-8B & framing=risk vs baseline & 10+30 & -1.00 & 2.4e-09 & 7.1e-09 \\
perturbation & pgg & Qwen3-8B & persona\_content=prosocial vs baseline & 10+30 & -1.00 & 2.4e-09 & 7.1e-09 \\
perturbation & pgg & Qwen3-8B & persona\_content=self\_interested vs b & 10+30 & -1.00 & 2.4e-09 & 7.1e-09 \\
perturbation & pgg & Qwen3-14B & framing=risk vs baseline & 10+30 & +1.00 & 2.4e-09 & 9.4e-09 \\
perturbation & pgg & Qwen3-14B & persona\_content=self\_interested vs b & 10+30 & -1.00 & 2.4e-09 & 9.4e-09 \\
\bottomrule
\end{tabular}}
\end{table*}

%% file: tables/si_contam.tex
\begin{table}[htbp]
\caption{Contamination outcomes away from the divergent-payoff variant, and what the models say when asked directly. C1 re-skins the game with a different cover story and action labels while preserving the payoffs; C2 additionally shifts the numbers. Values are the shift in the repeated dilemma's cooperation rate relative to the canonical baseline. The priming column is the shift when the paradigm is named inside the behavioural prompt. Recognition is asked in a separate session: whether the model reports knowing the paradigm, and whether it names it correctly. Recognition predicts none of the behavioural columns. R1-Distill completed a partial battery: it ran the C1 and divergent-payoff cells of the repeated dilemma but no C2 cell, so its C2 entry is absent rather than zero.}
\label{tab:si-contam}\small\setlength{\tabcolsep}{3pt}\centering
\begin{tabular}{@{}l r r r r r@{}}
\toprule
Model & C1 shift & C2 shift & Priming shift & Reports knowing & Names correctly \\
\midrule
Qwen3-32B & -0.05 & +0.05 & -0.03 & 100\% & 70\% \\
Qwen3-14B & -0.20 & -0.85 & +0.08 & 100\% & 70\% \\
Qwen3-8B & +0.00 & +0.00 & +0.00 & 100\% & 56\% \\
Qwen3-4B & +0.34 & +0.10 & +0.03 & 100\% & 57\% \\
Qwen2.5-14B & +0.00 & +0.00 & +0.00 & 80\% & 51\% \\
Qwen2.5-7B & +0.00 & -0.98 & +0.00 & 93\% & 55\% \\
Qwen1.5-7B & +0.90 & -0.04 & +0.06 & 19\% & 15\% \\
Gemma2-9B & -0.02 & -0.89 & +0.00 & 100\% & 55\% \\
Phi3-mini & +0.12 & +0.01 & +0.01 & 90\% & 43\% \\
Mistral-7B & +0.00 & -0.37 & +0.00 & 95\% & 53\% \\
Llama3.1-8B & +0.06 & -0.10 & +0.06 & 100\% & 42\% \\
R1-7B & -0.18 & -- & +0.15 & 90\% & 64\% \\
\bottomrule
\end{tabular}
\end{table}

%% file: tables/si_anchors.tex
\begin{table*}[htbp]
\caption{Provenance of every human anchor. Each value was checked against its primary source on 26 August 2026 and the shipped anchor file now records the citation and the verification date. One correction was made: two cells of the 11--20 distribution had been mis-transcribed, and all Wasserstein distances were recomputed against the published table. The public-goods trajectory is a stylised fact drawn from a survey and a meta-analysis rather than a single meta-analytic estimate, and its punishment value comes from one experiment; the manuscript states this rather than describing the anchor as a meta-analytic corridor.}
\label{tab:si-anchors}\scriptsize\setlength{\tabcolsep}{3pt}\centering
\begin{tabular}{@{}>{\raggedright\arraybackslash}p{0.215\textwidth}>{\raggedright\arraybackslash}p{0.185\textwidth}>{\raggedright\arraybackslash}p{0.215\textwidth}>{\raggedright\arraybackslash}p{0.215\textwidth}>{\raggedright\arraybackslash}p{0.120\textwidth}@{}}
\toprule
Environment & Anchor & Source & Evidence base & Provenance \\
\midrule
11--20 game & full choice distribution; $P(17)=.32$, $P(18)=.30$ & Arad \& Rubinstein (2012), AER 102(7), Table 1 & single experiment, $n=108$ & published table \\
Ultimatum & offer $.40$ of pie; rejection $.16$ & Oosterbeek et al.\ (2004), Exp.\ Econ.\ 7(2) & meta-analysis, 37 papers, 75 results & meta-analytic mean \\
Dictator & giving $.28$ of endowment & Engel (2011), Exp.\ Econ.\ 14 & meta-analysis, 328 treatments, 20{,}813 decisions & random-effects mean \\
Trust & sent $.50$; returned $.37$ of available & Johnson \& Mislin (2011), JEP 32(5) & meta-analysis, 162 replications, $>$23{,}000 participants & meta-analytic mean \\
Public goods & round 1 $.50$; round 10 $.15$; punishment $.55$ & Ledyard (1995) survey; Zelmer (2003); Fehr \& G\"achter (2000) & 27 studies, 711 groups; punishment from a single experiment & stylised fact, not a single estimate \\
Repeated PD & cooperation corridor $.20$--$.50$ & Mengel (2018), Econ.\ J.\ 128(616) & meta-study, 96 studies, $>$3{,}500 participants & range across studies \\
Naming / tipping & convention always forms; critical mass $.21$--$.25$ & Centola et al.\ (2018), Science 360 & 10 groups of 20 participants & single experimental study \\
\bottomrule
\end{tabular}
\end{table*}

%% file: tables/si_grid.tex
\begin{table*}[htbp]
\caption{Acceptance functions under the fixed offer schedule: rejection rate at each scheduled offer, 20 games per offer per seed, three seeds. In the canonical variant rejection pays nothing, so accepting every offer is payoff-maximising; in the divergent variant rejection pays 40, so rejecting below 40 is. The threshold is the largest offer still rejected in at least half of games. R1-Distill is the only model whose threshold falls where the outside option puts it.}
\label{tab:si-grid}\tiny\setlength{\tabcolsep}{3pt}\centering
\resizebox{\textwidth}{!}{%
\begin{tabular}{l l rrrrrrrrr r r}
\toprule
Model & Variant & 10 & 20 & 25 & 30 & 33 & 37 & 40 & 45 & 50 & Threshold & Consistency \\
\midrule
Qwen3-32B & canonical & 0.13 & 0.00 & 0.00 & 0.00 & 0.00 & 0.00 & 0.00 & 0.00 & 0.00 & -- & 0.99 \\
 & divergent & 1.00 & 0.53 & 0.67 & 0.68 & 0.00 & 0.03 & 0.05 & 0.00 & 0.00 & 30 & 0.61 \\
Qwen3-14B & canonical & 1.00 & 1.00 & 1.00 & 0.00 & 0.00 & 0.00 & 0.00 & 0.00 & 0.00 & 25 & 0.67 \\
 & divergent & 1.00 & 1.00 & 1.00 & 1.00 & 0.75 & 0.13 & 0.00 & 0.00 & 0.00 & 33 & 0.86 \\
Qwen3-8B & canonical & 0.00 & 0.00 & 0.00 & 0.00 & 0.00 & 0.00 & 0.00 & 0.00 & 0.00 & -- & 1.00 \\
 & divergent & 0.00 & 0.00 & 0.00 & 0.00 & 0.00 & 0.00 & 0.00 & 0.00 & 0.00 & -- & 0.25 \\
Qwen3-4B & canonical & 0.82 & 0.00 & 0.00 & 0.00 & 0.00 & 0.00 & 0.00 & 0.00 & 0.00 & 10 & 0.91 \\
 & divergent & 0.98 & 0.25 & 0.00 & 0.05 & 0.00 & 0.00 & 0.00 & 0.00 & 0.00 & 10 & 0.41 \\
Qwen2.5-14B & canonical & 1.00 & 1.00 & 0.25 & 0.00 & 0.72 & 0.00 & 0.00 & 0.00 & 0.00 & 33 & 0.67 \\
 & divergent & 0.62 & 0.00 & 0.20 & 0.00 & 0.00 & 0.00 & 0.00 & 0.00 & 0.00 & 10 & 0.35 \\
Qwen2.5-7B & canonical & 1.00 & 1.00 & 1.00 & 0.57 & 1.00 & 0.00 & 0.00 & 0.00 & 0.00 & 33 & 0.49 \\
 & divergent & 1.00 & 1.00 & 1.00 & 1.00 & 1.00 & 1.00 & 1.00 & 0.38 & 0.33 & 40 & 0.91 \\
Qwen1.5-7B & canonical & 1.00 & 1.00 & 1.00 & 1.00 & 1.00 & 1.00 & 1.00 & 1.00 & 1.00 & 50 & 0.00 \\
 & divergent & 0.70 & 0.73 & 1.00 & 0.63 & 1.00 & 1.00 & 1.00 & 1.00 & 0.65 & 50 & 0.68 \\
Gemma2-9B & canonical & 1.00 & 1.00 & 1.00 & 1.00 & 1.00 & 1.00 & 1.00 & 1.00 & 1.00 & 50 & 0.00 \\
 & divergent & 1.00 & 0.95 & 0.88 & 0.68 & 0.83 & 0.95 & 0.15 & 0.38 & 0.78 & 50 & 0.77 \\
Phi3-mini & canonical & 0.12 & 0.17 & 0.08 & 0.08 & 0.03 & 0.03 & 0.07 & 0.03 & 0.03 & -- & 0.93 \\
 & divergent & 0.35 & 0.40 & 0.08 & 0.20 & 0.05 & 0.20 & 0.15 & 0.05 & 0.08 & -- & 0.39 \\
Mistral-7B & canonical & 1.00 & 1.00 & 1.00 & 1.00 & 1.00 & 1.00 & 1.00 & 1.00 & 0.68 & 50 & 0.04 \\
 & divergent & 0.97 & 0.93 & 0.90 & 0.87 & 0.95 & 1.00 & 1.00 & 0.63 & 0.00 & 45 & 0.87 \\
Llama3.1-8B & canonical & 0.52 & 0.17 & 0.17 & 0.00 & 0.20 & 0.00 & 0.00 & 0.00 & 0.00 & 10 & 0.88 \\
 & divergent & 0.23 & 0.00 & 0.00 & 0.00 & 0.00 & 0.00 & 0.00 & 0.00 & 0.00 & -- & 0.28 \\
R1-7B & canonical & 0.32 & 0.15 & 0.28 & 0.25 & 0.32 & 0.43 & 0.17 & 0.27 & 0.05 & -- & 0.75 \\
 & divergent & 0.98 & 0.98 & 0.97 & 0.98 & 0.95 & 0.97 & 0.23 & 0.02 & 0.00 & 37 & 0.98 \\
\bottomrule
\end{tabular}}
\end{table*}

%% file: tables/si_shufpool.tex
\begin{table*}[htbp]
\caption{The shuffled-pool control. Each agent receives its own random permutation of the same ten names, and an unparsable reply falls back to a random name rather than the first listed one. Fixed-order convergence is the median over the 30 baseline runs; 50 is the earliest value the convergence rule can return. Coordination survives in every model and slows in all of them. Runs are listed separately by cap: Qwen2.5-7B reached the 600-interaction cap in all three capped runs and, uncapped, converged in only two of three. Every model still converges on the same name as under a fixed order except R1-Distill, which moves from A to E.}
\label{tab:si-shufpool}\scriptsize\setlength{\tabcolsep}{3pt}\centering
\resizebox{\textwidth}{!}{%
\begin{tabular}{l r r r r r l r l}
\toprule
Model & Fixed order & Cap & Runs & Converged & Median & Range & First-20 match & Name \\
\midrule
Qwen3-32B & 50 & 600 & 3 & 3 & 104 & 90--117 & 0.67 & A \\
Qwen3-14B & 50 & 600 & 3 & 3 & 103 & 100--237 & 0.42 & A \\
Qwen3-8B & 50 & 600 & 3 & 3 & 75 & 50--113 & 0.85 & A \\
Qwen3-4B & 50 & 600 & 3 & 3 & 129 & 104--202 & 0.58 & A \\
Qwen2.5-14B & 50 & 600 & 3 & 3 & 81 & 79--97 & 0.68 & A \\
Qwen2.5-7B & 50 & 600 & 3 & 0 & 600 & 600--600 & 0.17 & C/G \\
Qwen2.5-7B & 50 & 3,000 & 3 & 2 & 1214 & 684--3000 & 0.18 & C/G \\
Gemma2-9B & 130 & 600 & 3 & 3 & 250 & 164--272 & 0.30 & A \\
R1-7B & 169 & 3,000 & 5 & 5 & 156 & 103--231 & 0.25 & E \\
\bottomrule
\end{tabular}}
\end{table*}

%% file: tables/positioning.tex
\begin{table*}[t]
\centering
\caption{\textbf{Positioning against the closest prior work.} Rows name the benchmark or
platform where one exists and the study otherwise. Unlike every other table in this article, these
entries are editorial judgements about other authors' work rather than measurements: a mark records
only whether the cited work set that capability as an objective, not how well it succeeded. The
criteria are defined in the text above. \checkmark{} stated as an objective, with results reported
against it; (\checkmark) a partial or methodological precursor, present but not carried through to a
reported result; -- not an objective of the cited work, which is a statement about scope rather than
a criticism.}
\label{tab:positioning}
\scriptsize
\setlength{\tabcolsep}{3.2pt}
\resizebox{\textwidth}{!}{%
\begin{tabular}{p{3.0cm}ccccccccc}
\toprule
 & \shortstack{Interactive\\multi-agent} & \shortstack{Human-anchored\\(distribution)} & \shortstack{Perturbation\\audit} & \shortstack{Contamination\\audit (C3)} & \shortstack{$\geq$5 model\\families} & \shortstack{Scale\\axis} & \shortstack{Generation\\axis} & \shortstack{Claim\\certification} & \shortstack{Open exec.\\benchmark} \\
\midrule
\citet{park2024} & \checkmark & \checkmark & -- & -- & -- & -- & -- & -- & (\checkmark) \\
\citet{ashery2025} & \checkmark & (\checkmark) & (\checkmark) & (\checkmark) & (\checkmark) & -- & -- & -- & (\checkmark) \\
\citet{akata2025} & \checkmark & (\checkmark) & -- & -- & (\checkmark) & -- & -- & -- & -- \\
\citet{cui2025} & -- & \checkmark & -- & -- & -- & -- & -- & -- & (\checkmark) \\
\citet{gao2025pnas} & -- & \checkmark & (\checkmark) & (\checkmark) & \checkmark & -- & -- & -- & (\checkmark) \\
GTBench \citep{duan2024gtbench} & \checkmark & -- & -- & -- & \checkmark & (\checkmark) & -- & -- & \checkmark \\
GLEE \citep{shapira2024glee} & \checkmark & (\checkmark) & -- & -- & \checkmark & -- & -- & -- & \checkmark \\
$\gamma$-Bench \citep{huang2024gama} & \checkmark & -- & (\checkmark) & (\checkmark) & \checkmark & -- & -- & -- & \checkmark \\
M3-Bench \citep{m3bench2026} & \checkmark & (\checkmark) & -- & -- & \checkmark & -- & -- & -- & \checkmark \\
SimBench \citep{hu2025simbench} & -- & \checkmark & -- & -- & \checkmark & \checkmark & (\checkmark) & -- & \checkmark \\
TRAILS \citep{ye2026} & \checkmark & -- & \checkmark & (\checkmark) & (\checkmark) & -- & -- & (\checkmark) & (\checkmark) \\
PIMMUR \citep{zhou2025pimmur} & \checkmark & -- & (\checkmark) & (\checkmark) & (\checkmark) & -- & -- & (\checkmark) & -- \\
EASE \citep{sarangi2026} & \checkmark & -- & (\checkmark) & -- & (\checkmark) & -- & -- & -- & \checkmark \\
\citet{larooij2025} & -- & -- & (\checkmark) & (\checkmark) & -- & -- & -- & (\checkmark) & -- \\
\citet{barrie2025} & -- & -- & -- & (\checkmark) & -- & -- & -- & -- & -- \\
\citet{payne2026} & -- & (\checkmark) & -- & (\checkmark) & -- & -- & -- & -- & -- \\
\midrule
\textbf{SILICA (this work)} & \checkmark & \checkmark & \checkmark & \checkmark & \checkmark & \checkmark & \checkmark & \checkmark & \checkmark \\
\bottomrule
\end{tabular}}
\end{table*}